\documentclass[journal,draftcls,onecolumn,letterpaper,twoside,11pt]{IEEEtran}
\usepackage{color}

\usepackage[cmex10]{amsmath}
\usepackage{latexsym,amssymb,cite,bbm,epsf,amsthm,subfigure,algorithm,algorithmic}
\usepackage{graphicx}
\usepackage{enumerate}
\usepackage{mathrsfs}
\usepackage{bbm}
\usepackage{epstopdf}

\newcommand{\bc}{\begin{center}}
\newcommand{\ec}{\end{center}}
\newcommand{\be}{\begin{equation}}
\newcommand{\ee}{\end{equation}}
\newcommand{\ba}{\begin{array}}
\newcommand{\ea}{\end{array}}
\newcommand{\bea}{\begin{eqnarray}}
\newcommand{\eea}{\end{eqnarray}}
\newcommand{\bal}{\begin{align}}
\newcommand{\eal}{\end{align}}
\newcommand{\ei}{\end{itemize}}
\newcommand{\bi}{\begin{itemize}}
\newcommand{\bfi}{\begin{figure}}
\newcommand{\efi}{\end{figure}}
\newcommand{\MB}{\left[\begin{array}}
\newcommand{\ME}{\end{array}\right]}
\newcommand{\nn}{\nonumber}

\newtheorem{thm}{Theorem}

\newtheorem{lem}{Lemma}
\newtheorem{pro}{Proposition}

\newcommand{\Exp}{\mathsf{E}}
\newcommand{\Pro}{\mathsf{P}}
\newcommand{\Hyp}{\mathsf{H}}

\newcommand{\cT}{\mathcal{T}}

\newcommand{\cN}{\mathcal{N}}
\newcommand{\cY}{\mathcal{Y}}
\newcommand{\cC}{\mathcal{C}}

\newcommand{\bN}{\mathbb{N}}

\newcommand{\ind}[1]{\mathbbm{1}_{\{#1\}}}   %indicator

\newcommand{\ignore}[1]{{}}

\makeatletter
\renewcommand\subsubsection{\@startsection{subsubsection}{3}{5mm}
{-0.3\baselineskip}
{0.1\baselineskip}
{\normalfont\normalsize}}
\makeatother

\begin{document}

\title{Channel-aware Decentralized Detection via Level-triggered Sampling}

\author{Yasin~Yilmaz\IEEEauthorrefmark{2}\footnote{\IEEEauthorrefmark{2}Electrical Engineering Department, Columbia University, New York, NY 10027.},\;\;
        George V. Moustakides\IEEEauthorrefmark{3}\footnote{\IEEEauthorrefmark{3}Dept. of Electrical \& Computer Engineering, University of Patras, 26500 Rion, Greece.},\;\;
        and \, Xiaodong Wang\IEEEauthorrefmark{2}}
\ignore{\thanks{Y. Yilmaz and X. Wang are with the Department
of Electrical Engineering, Columbia University, New York,
NY, 10027 USA.}% <-this % stops a space
\thanks{G. Moustakides is with the Dept. of Electrical \& Computer Engineering, University of Patras, 26500 Rion, Greece.}}% <-this % stops a space

\maketitle

\begin{abstract}

We consider decentralized detection through distributed sensors that perform level-triggered sampling and communicate with a fusion center (FC) via noisy channels.  Each sensor computes its local log-likelihood ratio (LLR), samples it using the level-triggered sampling mechanism, and at each sampling instant transmits a single bit to the FC. Upon receiving a bit from a sensor, the
FC updates the global LLR and performs a sequential probability ratio test (SPRT) step. We derive the fusion rules under various types of channels. We further provide an asymptotic analysis on the average decision delay for the proposed channel-aware scheme, and show that the asymptotic decision delay is
characterized by a Kullback-Leibler information number. The delay analysis facilitates the choice of the appropriate signaling schemes under different channel types for sending the 1-bit information from the sensors to the FC.

\end{abstract}


\ignore{
}

%\newpage
\section{Introduction}

We consider the problem of binary decentralized detection where a number of distributed sensors, under bandwidth constraints, communicate with a fusion center (FC) which is responsible for making the final decision.
In \cite{Tenney81} it was shown that under a fixed fusion rule, with two sensors each transmitting one bit information to the FC, the optimum local decision rule is a likelihood ratio test (LRT) under the Bayesian criterion.
Later, in \cite{Chair86} and \cite{Thomo87} it was shown that the optimum fusion rule at the FC is also an LRT under the Bayesian and the Neyman-Pearson criteria, respectively.
It was further shown in \cite{Tsitsiklis88} that as the number of sensors tends to infinity it is asymptotically optimal to have all sensors perform an identical LRT.
The case where sensors observe correlated signals was also considered, e.g., \cite{Aalo89}, \cite{Willett00}.

Most works on decentralized detection, including the above mentioned, treat the fixed-sample-size approach where each sensor collects a fixed number of samples and the FC makes its final decision at a fixed time. There is also a significant volume of literature that considers the sequential detection approach where both, the sensor local decision times and the FC global decision time are random, e.g., \cite{Veer93,Mei08,Chaud09,Hussain94,Fellouris11,Yilmaz11}. Regarding references \cite{Hussain94,Fellouris11,Yilmaz11} we should mention that they use, both locally and globally, the sequential probability ratio test (SPRT), which is known to be optimal for i.i.d. observations in terms of minimizing the average sample number (decision delay) among all sequential tests satisfying the same error probability constraints \cite{Wald48}. SPRT has been shown in \cite[Page 109]{Poor} to asymptotically require, on average, four times less samples (for Gaussian signals) to reach a decision than the best fixed-sample-size test, for the same level of confidence.
Relaxing the one-bit messaging constraint, the optimality of the likelihood ratio quantization is established in \cite{Warren89}. Data fusion (multi-bit messaging) is known to be much more powerful than decision fusion (one-bit messaging) \cite{Chaud12}, albeit it consumes higher bandwith. Moreover,  the recently proposed sequential detection schemes  based on level-triggered sampling in \cite{Fellouris11} and \cite{Yilmaz11} are as powerful as data-fusion techniques, and at the same time they are as simple and bandwidth-efficient as decision-fusion techniques.

Besides having noisy observations at sensors, in practice the channels between sensors and the FC are noisy. The conventional approach to decentralized detection ignores the latter, i.e., assumes ideal transmission channels, and addresses only the first source of uncertainty, e.g., \cite{Tenney81}, \cite{Fellouris11}. Adopting the conventional approach to the noisy channel case yields a two-step solution. First, a communication block is employed at the FC to recover the transmitted information bits from sensors, and then a signal processing block applies a fusion rule to the recovered bits to make a final decision. Such an independent block structure causes performance loss due to the data processing inequality \cite{Chen06}. To obtain the optimum performance the FC should process the received signal in a channel-aware manner \cite{Chamberland03}, \cite{Liu06}.
Most works assume parallel channels between sensors and the FC, e.g., \cite{Chen04}, \cite{Niu06}. Other topologies such as serial \cite{Bahceci05} and multiple-access channels (MAC) \cite{Tepedelen11} have also been considered. In \cite{Ahmadi11}  a scheme is proposed that adaptively switches between serial and parallel topologies.

In this paper, we design and analyze channel-aware sequential decentralized detection schemes based on level-triggered sampling, under different types of discrete and continuous noisy channels. In particular, we first derive channel-aware sequential detection schemes based on level-triggered sampling. We then present an information theoretic framework to analyze the decision delay performance of the proposed schemes based on which we provide an asymptotic analysis on the decision delays under various types of channels. Based on the expressions  of the asymptotic decision delays, we also consider appropriate signaling schemes under different continuous channels to minimize the asymptotic delays.

The remainder of the paper is organized as follows. In Section \ref{sec:back}, we describe the general structure of the decentralized detection approach based on level-triggered sampling with noisy channels between sensors and the FC. In Section \ref{sec:Alg}, we derive channel-aware fusion rules at the FC for various types of channels. Next, we provide analyses on the decision delay performance for ideal channel and noisy channels  in Section \ref{sec:perf} and Section \ref{sec:Per_noi}, respectively. In Section \ref{sec:Unrel_det}, we discuss the issue of unreliable detection of the sensor sampling times by the FC. Simulation results are provided in Section \ref{sec:sim}. Finally, Section \ref{sec:conc} concludes the paper.

\section{System Descriptions}
\label{sec:back}

Consider a wireless sensor network consisting of $K$ sensors each of which observes a Nyquist-rate sampled discrete-time signal $\{y^k_t, t\in\bN\}, k=1,\ldots,K$. Each sensor $k$ computes the log-likelihood ratio (LLR) $\{ L_t^k, t \in \bN\}$ of the signal it observes, samples the LLR sequence using the {\em level-triggered sampling}, and then sends the LLR samples to the fusion center (FC). The FC then combines the local LLR information from all sensors, and decides between two hypotheses, $\Hyp_0$ and $\Hyp_1$, in a sequential manner.

\begin{figure}[t]
\centering
\includegraphics[scale=0.8]{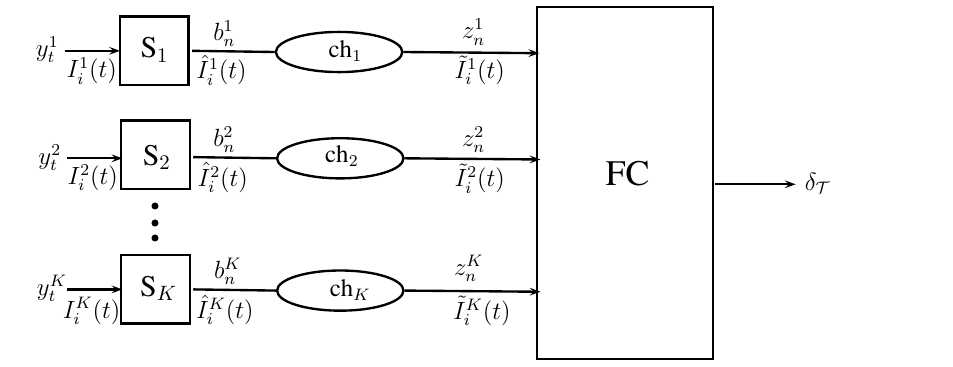}
\caption{A wireless sensor network with $K$ sensors $S_1,\ldots,S_K$, and a fusion center (FC). Sensors process their observations $\{y^k_t\}$, and transmits information bits $\{b^k_n\}$. Then, the FC, receiving $\{z^k_n\}$ through wireless channels, makes a detection decision $\delta_{\tilde{\cT}}$. $I^k_i(t),~\hat{I}^k_i(t),~\tilde{I}^k_i(t)$ are the observed, transmitted and received information entities respectively, which will be defined in Section \ref{sec:perf}.}
\label{fig:wsn}
\end{figure}

Observations collected at the same sensor, $\{y^k_t\}_t$, are assumed to be i.i.d., and in addition observations collected at different sensors, $\{y^k_t\}_k$, are assumed to be independent. Hence, the local LLR at the $k$-th sensor, $L^k_t$, and the global LLR, $L_t$, are computed as
\begin{align}
\label{eq:locLLR}
L_t^k \triangleq \log \frac{f^k_1(y_1^k,\ldots,y_t^k)}{f^k_0(y_1^k,\ldots,y_t^k)} = L^k_{t-1} + l^k_t=\sum_{n=1}^t l^k_n, \ \ \ \
\mbox{and} \ \ \ \ L_t = \sum_{k=1}^K L_t^k,
\end{align}
respectively, where $l^k_t \triangleq \log\frac{f^k_1(y^k_t)}{f^k_0(y^k_t)}$ is the LLR of the sample $y^k_t$ received at the $k$-th sensor at time $t$; $f^k_i, ~i=0,1$, is the probability density function (pdf) of the received signal by the $k$-th sensor under $\Hyp_i$. The $k$-th sensor samples $L^k_t$ via the level-triggered sampling at a sequence of random sampling times $\{t^k_n\}_n$ that are dictated by $L^k_t$ itself. Specifically, the $n$-th sample is taken from $L^k_t$ whenever the accumulated LLR $L^k_t-L^k_{t^k_{n-1}}$, since the last sampling time $t^k_{n-1}$ exceeds a constant $\Delta$ in absolute value, i.e.,
\be
\label{eq:LSsamptime}
t_n^k \triangleq \inf \left\{t > t^k_{n-1}: L^k_t - L^k_{t^k_{n-1}} \not\in(-\Delta,\Delta)\right\},~t^k_0=0,~L^k_0=0.
\ee
Let $\lambda^k_n$ denote the accumulated LLR during the $n$-th inter-sampling interval, $(t^k_{n-1},t^k_n]$, i.e.,
\be	\label{eq:loc_LLR}
\lambda^k_n \triangleq \sum_{t=t^k_{n-1}+1}^{t^k_n} l^k_t = L^k_{t^k_n}-L^k_{t^k_{n-1}}.
\ee
Immediately after sampling at $t^k_n$, as shown in Fig. \ref{fig:wsn}, an information bit $b^k_n$ indicating the threshold crossed by $\lambda^k_n$ is transmitted to the FC, i.e.,
\be	\label{eq:inf_bit}
b^k_n \triangleq \text{sign}(\lambda^k_n).
\ee

Note that each sensor, in fact, implements a local SPRT [cf. \eqref{eq:ideal1}, \eqref{eq:stop_time}], with thresholds $\Delta$ and $-\Delta$ within each sampling interval. At sensor $k$ the $n$-th local SPRT starts at time $t^k_{n-1}$ and ends at time $t^k_n$ when the local test statistic $\lambda^k_n$ exceeds either $\Delta$ or $-\Delta$. This local hypothesis testing produces a local decision represented by the information bit $b^k_n$, and induces local error probabilities $\alpha_k$ and $\beta_k$ which are given by
\be	\label{eq:loc_err}
\alpha_k \triangleq \Pro_0(b^k_n=1), \ \ \ \
\mbox{and} \ \ \ \  \beta_k \triangleq \Pro_1(b^k_n=-1)
\ee
respectively, where $\Pro_i(\cdot),~i=0,1$, denotes the probability under $\Hyp_i$.

Let us now analyze the signals at the FC. Denote the received signal at the FC corresponding to $b_n^k$ as $z_n^k$ [cf. Fig. \ref{fig:wsn}].  The FC then computes the LLR $\tilde \lambda_n^k$ of each received signal
and approximates the global LLR $L_t$ as
\be
\label{eq:rec_up}
\tilde{L}_t  \triangleq \sum_{k=1}^K \sum_{n=1}^{N^k_t} \tilde{\lambda}^k_n ~~~~ \mbox{with} ~~~~ \tilde \lambda_n^k \triangleq \log \frac {p^k_1(z^k_n)} {p^k_0(z^k_n)},
\ee
where $N^k_t$ is the total number of LLR messages the $k$-th sensor has transmitted up to time $t$, and $p^k_i(\cdot),~i=0,1$, is the pdf of $z^k_n$ under $\Hyp_i$. In fact, the FC recursively updates $\tilde L_t$ whenever it receives an LLR message from any sensor. In particular, suppose that the $m$-th LLR message $\tilde \lambda_m$ from any sensor is received at time $t_m$. Then at $t_m$, the FC first updates the global LLR as
\be
\label{eq:global_update}
\tilde L_{t_m} = \tilde L_{t_{m-1}} + \tilde \lambda_m.
\ee
It then performs an SPRT step by comparing $\tilde L _{t_m}$ with two thresholds $\tilde A$ and $-\tilde B$, and applying the following decision rule
\be     \label{eq:ideal1}
\delta_{t_m} \triangleq \left\{  \begin{array}{ll}
\Hyp_1, & \mbox{if $\tilde{L}_{t_m}\geq \tilde A$}, \\
\Hyp_0, & \mbox{if $\tilde{L}_{t_m}\leq -\tilde B$}, \\
\mbox{continue to receive LLR messages}, & \mbox{if $\tilde{L}_{t_m}\in (-\tilde B, \tilde A)$}.
\end{array} \right.
\ee
The thresholds ($\tilde A,\tilde B>0$) are selected to satisfy the error probability constraints  $\Pro_0(\delta_{\tilde{\cT}}=\Hyp_1) \leq \alpha$ and $\Pro_1(\delta_{\tilde{\cT}}=\Hyp_0) \leq \beta$ with equalities, where $\alpha,\beta$ are target error probability bounds, and
\be
\label{eq:stop_time}
\tilde{\cT} \triangleq \inf\{ t>0: \tilde{L}_t \not\in (-\tilde B,\tilde A) \}
\ee
is the decision delay.

With ideal channels between sensors and the FC, we have $z^k_n=b^k_n$, so from (\ref{eq:loc_err}) we can write the local LLR $\tilde{\lambda}^k_n=\hat{\lambda}^k_n$, where
\be 	\label{eq:ideal2}
\hat{\lambda}^k_n \triangleq \left\{  \begin{array}{ll}
\log \frac {\Pro_1(b^k_n=1)}  {\Pro_0(b^k_n=1)} = \log \frac{1-\beta_k} {\alpha_k} \geq \Delta, & \mbox{if $b^k_n=1$}, \\
\log \frac {\Pro_1(b^k_n=-1)}  {\Pro_0(b^k_n=-1)} = \log \frac{\beta_k} {1-\alpha_k} \leq -\Delta, & \mbox{if $b^k_n=-1$}
\end{array} \right.
\ee
where the inequalities can be easily obtained by applying a change of measure. For example, to show the first one, we have $\alpha_k = \Pro_0(\lambda^k_n \geq \Delta) = \Exp_0[\ind{\lambda^k_n \geq \Delta}]$ where $\Exp_i[\cdot]$ is the expectation under $\Hyp_i, i=0,1$ and $\ind{\cdot}$ is the indicator function.
Noting that $e^{-\lambda^k_n}=\frac{f^k_0(y^k_{t^k_{n-1}+1},\ldots,y^k_{t^k_n})} {f^k_1(y^k_{t^k_{n-1}+1},\ldots,y^k_{t^k_n})}$, we can write
\be
\alpha_k = \Exp_1[ e^{-\lambda^k_n} \ind{\lambda^k_n \geq \Delta}] \leq e^{-\Delta} \Exp_1[\ind{\lambda^k_n \geq \Delta}] = e^{-\Delta} \Pro_1(\lambda^k_n \geq \Delta) = e^{-\Delta} (1-\beta_k). \nn
\ee
Note that for the case of continuous-time and continuous-path observations at sensors, the inequalities in (\ref{eq:ideal2}) become equalities as the local LLR sampled at a sensor [cf. \eqref{eq:locLLR}] is now a continuous-time and continuous-path process. This suggests that the accumulated LLR during any inter-sampling interval [cf. \eqref{eq:loc_LLR}] due to continuity of its paths will hit exactly the local thresholds $\pm\Delta$. Therefore, from Wald's analysis for SPRT $\alpha_k=\beta_k=\frac{1}{e^{\Delta}+1}$ \cite{Wald47}; hence a transmitted bit fully represents the LLR accumulated in the corresponding inter-sampling interval. Accordingly, the FC at sampling times exactly recovers the values of LLR processes observed by sensors \cite{Fellouris11}.

When sensors observe discrete-time signals, due to randomly over(under)shooting the local thresholds, $\lambda^k_n$ in (\ref{eq:loc_LLR}) is a random variable which is in absolute value greater than $\Delta$. However, $\hat{\lambda}^k_n$ in (\ref{eq:ideal2}) is a fixed value that is also greater than $\Delta$ in absolute value.
While in continuous-time the FC fully recovers the LLR accumulated in an inter-sampling interval by using only the received bit, in discrete-time this is not possible. In order to ameliorate this problem, in  \cite{Fellouris11} it is assumed that the local error probabilities $\{\alpha_k, \beta_k\}$ are available to the FC; and therefore the LLR of $z_n^k$, i.e., $\hat \lambda_n^k$, can be obtained; while in \cite{Yilmaz11} the overshoot is quantized by using extra bits in addition to $b^k_n$. Nevertheless, neither method enables the FC to fully recover $\lambda^k_n$ unless an infinite number of bits is used.
In this paper, to simplify the performance analysis, we will assume, as in \cite{Fellouris11}, that the local error probabilities $\alpha_k,\beta_k,~k=1,\ldots,K$ are available at the FC in order to compute the LLR $\tilde \lambda_n^k$  of the received signals.
Moreover, for the case of ideal channels, we use the $A$ and $-B$ to denote the thresholds in (\ref{eq:ideal1}), i.e., $\tilde A=A, \tilde B=B$, and use $\cT$ to denote the decision delay in (\ref{eq:stop_time}), i.e.,  $\tilde{\cT}=\cT$.

In the case of noisy channels, the received signal $z^k_n$ is not always identical to the transmitted bit $b^k_n$, and thus the LLR $\tilde{\lambda}^k_n$ of $z^k_n$ can be different from $\hat{\lambda}^k_n$ of $b^k_n$ given in (\ref{eq:ideal2}). In the next section, we consider some popular channel models and give the corresponding expressions for $\tilde\lambda_n^k$.

\section{Channel-aware Fusion Rules}
\label{sec:Alg}

In computing the LLR $\tilde\lambda_n^k$ of the received signal $z_n^k$, we will make use of the local sensor error probabilities $\alpha_k, \beta_k$, and the channel parameters that characterize the statistical property of the channel. One subtle issue is that since the sensors asynchronously sample and transmit the local LLR, in the presence of noisy channels, the FC needs to first reliably detect the sampling time in order to update the global LLR. In this section we assume that the sampling time is reliably detected and focus on deriving the fusion rule at the FC. In Section \ref{sec:Unrel_det}, we will discuss the issue of sampling time detection.

\subsection{Binary Erasure Channels (BEC)}
\label{sec:bec_alg}

Consider binary erasure channels between sensors and the FC with erasure probabilities $\epsilon_k,~k=1,\ldots,K$. Under BEC, a transmitted bit $b^k_n$ is lost with probability $\epsilon_k$, and correctly received at the FC, i.e., $z^k_n=b^k_n$, with probability $1-\epsilon_k$. Then the LLR of $z^k_n$ is given by
\be 	\label{eq:bec1}
\tilde{\lambda}^k_n = \left\{  \begin{array}{ll}
\log \frac {\Pro_1(z^k_n=1)}  {\Pro_0(z^k_n=1)} = \log \frac{1-\beta_k} {\alpha_k}, & \mbox{if $z^k_n=1$}, \\
\log \frac {\Pro_1(z^k_n=-1)}  {\Pro_0(z^k_n=-1)} = \log \frac{\beta_k} {1-\alpha_k}, & \mbox{if $z^k_n=-1$}.
\end{array} \right.
\ee

Note that under BEC the channel parameter $\epsilon_k$ is not needed when computing the LLR $\tilde\lambda_n^k$.
Note also that in this case, a received bit bears the same amount of LLR information %, $\pm\Delta$,
as in the ideal channel case, although a transmitted bit is not always received.
Hence, the channel-aware approach coincides with the conventional approach which relies solely on the received signal.
Although the LLR updates in (\ref{eq:ideal2}) and (\ref{eq:bec1}) are identical, the fusion rules under BEC and ideal channels are not. This is because the thresholds $\tilde{A}$ and $-\tilde{B}$ of BEC, due to the information loss, are in general different from the thresholds $A$ and $-B$ of the ideal channel case.

\subsection{Binary Symmetric Channels (BSC)}
\label{sec:bsc_alg}

Next, we consider binary symmetric channels with crossover probabilities $\epsilon_k$ between sensors  and the FC. Under BSC, the transmitted bit $b^k_n$ is flipped, i.e., $z^k_n=-b^k_n$, with probability $\epsilon_k$, and it is correctly received, i.e., $z^k_n=b^k_n$, with probability $1-\epsilon_k$.
The LLR of $z^k_n$ can be computed as
\begin{align}
\tilde{\lambda}^k_n(z^k_n=1) &= \log \frac{ \Pro_1(z^k_n=1 | b^k_n=1) \Pro_1(b^k_n=1) + \Pro_1(z^k_n=1 | b^k_n=-1) \Pro_1(b^k_n=-1) }  { \Pro_0(z^k_n=1 | b^k_n=1) \Pro_0(b^k_n=1) + \Pro_0(z^k_n=1 | b^k_n=-1) \Pro_0(b^k_n=-1) } \nn\\
&= \log \frac{(1-\epsilon_k)(1-\beta_k)+\epsilon_k \beta_k} {(1-\epsilon_k) \alpha_k+\epsilon_k (1-\alpha_k)} = \log \frac{1- \overbrace{[(1-2\epsilon_k )\beta_k+\epsilon_k ]}^{\hat{\beta}_k}} {\underbrace{(1-2\epsilon_k )\alpha_k+\epsilon_k}_{\hat{\alpha}_k}}
\label{eq:bsc2}
\end{align}
where $\hat{\alpha}_k$ and $\hat{\beta}_k$ are the effective local error probabilities at the FC under BSC.
Similarly we can write
\be
\label{eq:bsc3}
\tilde{\lambda}^k_n(z^k_n=-1) = \log \frac{\hat{\beta}_k}{1-\hat{\alpha}_k}. %\leq -\hat{\Delta}_k.
\ee

Note that $\hat{\alpha}_k > \alpha_k,~\hat{\beta}_k > \beta_k$ if $\alpha_k<0.5,~\beta_k<0.5,~\forall k$, which we assume true for $\Delta>0$. Thus, we have $|\tilde{\lambda}^k_{n,BSC}|<|\tilde{\lambda}^k_{n,BEC}|$ from which we expect the performance loss under BSC to be higher than the one under BEC. The numerical results provided in Section \ref{sec:bsc_per} will illustrate this claim.
Finally, note also that, unlike the BEC case, under BSC the FC needs to know the channel parameters $\{\epsilon_k\}$ to operate in a channel-aware manner.

\subsection{Additive White Gaussian Noise (AWGN) Channels}
\label{sec:awgn_alg}

Now, assume that the channel between each sensor and the FC is an AWGN channel. The received signal at the FC is given by
\be
\label{eq:sigmodel_awgn}
z_n^k = h^k_n x_n^k + w_n^k
\ee
where $h^k_n=h_k, \forall k,n$, is a known constant complex channel gain;
$w_n^k \sim {\mathcal N}_c (0, \sigma_k^2)$;
$x_n^k$ is the transmitted signal at sampling time $t_n^k$ given by
\be
\label{eq:tx_sig}
x_n^k = \left\{ \begin{array}{ll}
        a, & \mbox{if $\lambda_n^k \geq \Delta$}, \\
        b, & \mbox{if $\lambda_n^k \leq -\Delta$}.
        \end{array} \right.
\ee
where the transmission levels $a$ and $b$ are complex in general.

The distribution of the received signal is then  $z_n^k \sim {\cal N}_c (h_k x_n^k, \sigma_k^2)$. The LLR of $z_n^k$ is given by
\begin{align}
\label{eq:LLR_awgn}
\tilde\lambda^k_n =& \log \frac {p_k(z_n^k | x_n^k=a) \Pro_1(x_n^k=a) + p_k(z_n | x_n^k=b) \Pro_1(x_n^k=b)}
{p_k(z_n^k | x_n^k=a) \Pro_0(x_n^k=a) + p_k(z_n^k | x_n^k=b) \Pro_0(x_n^k=b)} \nn\\
=& \log \frac { (1-\beta_k) \exp(-c^k_n) + \beta_k \exp(-d^k_n) } { \alpha_k \exp(-c^k_n) + (1-\alpha_k) \exp(-d^k_n) },
\end{align}
where $c^k_n \triangleq \frac{|z^k_n-h_k a|^2}{\sigma_k^2}$ and $d^k_n \triangleq \frac{|z^k_n-h_k b|^2}{\sigma_k^2}$.

\subsection{Rayleigh Fading Channels}
\label{sec:ray_alg}

If a Rayleigh fading channel is assumed between each sensor and the FC, the received signal model is also given by (\ref{eq:sigmodel_awgn})-(\ref{eq:tx_sig}), but with $h^k_n \sim \mathcal{N}_c (0,\sigma^2_{h,k})$. We then have $z_n^k \sim {\cal N}_c (0,|x^k_n|^2\sigma_{h,k}^2+\sigma_k^2)$;
and accordingly, similar to (\ref{eq:LLR_awgn}), $\tilde\lambda^k_n$ is written as
\be
\label{eq:LLR_ray}
\tilde \lambda^k_n = \log
\frac {  \frac{1-\beta_k}{\sigma_{a,k}^2} \exp (-c^k_n) + \frac{\beta_k}{\sigma_{b,k}^2} \exp (-d^k_n) }
{ \frac{\alpha_k}{\sigma_{a,k}^2} \exp (-c^k_n) + \frac{1-\alpha_k}{\sigma_{b,k}^2} \exp (-d^k_n) }
\ee
where $\sigma_{a,k}^2 \triangleq |a|^2\sigma_{h,k}^2+\sigma_k^2$, $\sigma_{b,k}^2 \triangleq |b|^2\sigma_{h,k}^2+\sigma_k^2$, $c^k_n \triangleq \frac{|z^k_n|^2}{\sigma_{a,k}^2}$ and $d^k_n \triangleq \frac{|z^k_n|^2}{\sigma_{b,k}^2}$.

\subsection{Rician Fading Channels}
\label{sec:ric_alg}

For Rician fading channels, we have $h^k_n \sim \mathcal{N}_c (\mu_k,\sigma^2_{h,k})$ in (\ref{eq:sigmodel_awgn}), and hence $z_n^k \sim {\cal N}_c (\mu_k x^k_n, |x^k_n|^2\sigma_{h,k}^2+\sigma_k^2)$.
Using $\sigma_{a,k}^2$ and $\sigma_{b,k}^2$ as defined in the Rayleigh fading case, and defining $c^k_n \triangleq \frac{|z^k_n-\mu_k a|^2}{\sigma_{a,k}^2}$, $d^k_n \triangleq \frac{|z^k_n-\mu_k b|^2}{\sigma_{b,k}^2}$ we can write $\tilde \lambda^k_n$ as in (\ref{eq:LLR_ray}).

\section{Performance Analysis for Ideal Channels}
\label{sec:perf}

In this section, we first find the non-asymptotical expression for the average decision delay $\Exp_i[\cT]$, and then
provide an asymptotic analysis on it as the error probability bounds $\alpha,\beta\to 0$.
Before proceeding to the analysis, let us define some information entities which will be used throughout this and next sections.

\subsection{Information Entities}	\label{sec:info}

Note that the expectation of an LLR corresponds to a \emph{Kullback-Leibler (KL) information} entity. For instance,
\be
I^k_1(t) \triangleq \Exp_1\left[ \log\frac{f^k_1(y^k_1,\ldots,y^k_t)}{f^k_0(y^k_1,\ldots,y^k_t)} \right] = \Exp_1[L^k_t],
~~\text{and}~~
I^k_0(t) \triangleq \Exp_0\left[ \log\frac{f^k_0(y^k_1,\ldots,y^k_t)}{f^k_1(y^k_1,\ldots,y^k_t)} \right] = -\Exp_0[L^k_t]
\ee
 are the KL divergences of the local LLR sequence $\{L^k_t\}_t$ under $\Hyp_1$ and $\Hyp_0$, respectively. Similarly
 \begin{align}
 \begin{split}
 \hat{I}^k_1(t) &\triangleq \Exp_1\left[ \log\frac{p^k_1(b^k_1,\ldots,b^k_{N^k_t})}{p^k_0(b^k_1,\ldots,b^k_{N^k_t})} \right] = \Exp_1[\hat{L}^k_t] ~~\text{,}~~ \hat{I}^k_0(t) \triangleq -\Exp_0[\hat{L}^k_t] \\
\tilde{I}^k_1(t) &\triangleq \Exp_1\left[ \log\frac{p^k_1(z^k_1,\ldots,z^k_{N^k_t})}{p^k_0(z^k_1,\ldots,z^k_{N^k_t})} \right] = \Exp_1[\tilde{L}^k_t] ~~\text{,}~~ \tilde{I}^k_0(t) \triangleq -\Exp_0[\tilde{L}^k_t]
\end{split}
\end{align}
are the KL divergences of the local LLR sequences $\{\hat{L}^k_t\}_t$ and $\{\tilde{L}^k_t\}_t$ respectively. Define also $I_i(t) \triangleq \sum_{k=1}^K I^k_i(t)$, $\hat{I}_i(t) \triangleq \sum_{k=1}^K \hat{I}^k_i(t)$, and $\tilde{I}_i(t) \triangleq \sum_{k=1}^K \tilde{I}^k_i(t)$ as the KL divergences of the global LLR sequences $\{L_t\}$, $\{\hat{L}_t\}$, and $\{\tilde{L}_t\}$ respectively.

In particular, we have
\be
I^k_1(1)=\Exp_1\left[ \log\frac{f^k_1(y^k_1)}{f^k_0(y^k_1)} \right]=\Exp_1[l^k_1],~\text{and}~I^k_0(1)=\Exp_0\left[ \log\frac{f^k_0(y^k_1)}{f^k_1(y^k_1)} \right]=-\Exp_0[l^k_1]
\ee
as the KL information numbers of the LLR sequence $\{l^k_t\}$; and $I_i(1) \triangleq \sum_{k=1}^K I^k_i(1),~i=0,1$ are those of the global LLR sequence $\{l_t\}$. Moreover,
\begin{align}
\begin{split}
I^k_1(t^k_1)=\Exp_1\left[ \log\frac{f^k_1(y^k_1,\ldots,y^k_{t^k_1})}{f^k_0(y^k_1,\ldots,y^k_{t^k_1})} \right]=&\Exp_1[\lambda^k_1],
~\hat{I}^k_1(t^k_1)=\Exp_1\left[ \log\frac{p^k_1(b^k_1)}{p^k_0(b^k_1)} \right]=\Exp_1[\hat{\lambda}^k_1],\\
\text{and}~\tilde{I}^k_1(t^k_1)=&\Exp_1\left[ \log\frac{p^k_1(z^k_1)}{p^k_0(z^k_1)} \right]=\Exp_1[\tilde{\lambda}^k_1]
\end{split}
\end{align}
are the KL information numbers of the local LLR sequences $\{\lambda^k_n\}$, $\{\hat{\lambda}^k_n\}$, and $\{\tilde{\lambda}^k_n\}$, respectively, under $\Hyp_1$. Likewise, we have $I^k_0(t^k_1)=-\Exp_0[\lambda^k_n]$, $\hat{I}^k_0(t^k_1)=-\Exp_0[\hat{\lambda}^k_n]$, and $\tilde{I}^k_0(t^k_1)=-\Exp_0[\tilde{\lambda}^k_n]$ under $\Hyp_0$.
To summarize, $I^k_i(t)$, $\hat{I}^k_i(t)$, and $\tilde{I}^k_i(t)$ are respectively the observed (at sensor $k$), transmitted (by sensor $k$), and received (by the FC) KL information entities as illustrated in Fig. \ref{fig:wsn}.

Next we define the following information ratios,
\be
\hat{\eta}^k_i \triangleq \frac{\hat{I}^k_i(t^k_1)}{I^k_i(t^k_1)},~\text{and}~\tilde{\eta}^k_i \triangleq \frac{\tilde{I}^k_i(t^k_1)}{I^k_i(t^k_1)},
\ee
which represent how efficiently information is transmitted from sensor $k$ and received by the FC, respectively. Due to the data processing inequality, we have $0\leq \hat{\eta}^k_i, \tilde{\eta}^k_i \leq1$, for $i=0,1$ and $k=1,\ldots,K$.
We further define
\be
\hat{I}_i(1) \triangleq \sum_{k=1}^K \hat{\eta}^k_i I^k_i(1) = \sum_{k=1}^K \hat{I}^k_i(1),~\text{and}~\tilde{I}_i(1) \triangleq \sum_{k=1}^K \tilde{\eta}^k_i I^k_i(1) = \sum_{k=1}^K \tilde{I}^k_i(1)
\ee
as the effective transmitted and received values corresponding to the KL information $I_i(1)$, respectively. Note that $\hat{I}_i(1)$ and $\tilde{I}_i(1)$ are not real KL
information numbers, but projections of $I_i(1)$ onto the filtrations generated by the transmitted, (i.e., $\{b^k_n\}$), and received, (i.e., $\{z^k_n\}$), signal sequences, respectively.
This is because sensors do not transmit and the FC does not receive the LLR of a single observation, but instead they transmit and it receives the LLR messages of several observations. Hence, we cannot have the KL information for single observations at the two ends of the communication channel, but we can define hypothetical KL information to serve analysis purposes. In fact, the hypothetical information numbers $\hat{I}_i(1)$ and $\tilde{I}_i(1)$, defined using the information ratios $\hat{\eta}^k_i$ and $\tilde{\eta}^k_i$, are crucial for our analysis as will be seen in the following sections.

The KL information $I^k_i(1)$ of a sensor whose information ratio, $\tilde{\eta}^k_i$, is high and close to $1$ is well projected to the FC. Conversely, $I^k_i(1)$ of a sensor which undergoes high information loss is poorly projected to the FC. Note that there are two sources of information loss for sensors, namely, the overshoot effect due to having discrete-time observations and noisy transmission channels. The latter appears only in $\tilde{\eta}^k_i$, whereas the former appears in both $\hat{\eta}^k_i$ and $\tilde{\eta}^k_i$.
In general with discrete-time observations at sensors we have $\hat{I}_i(1)\not=I_i(1)$ and $\tilde{I}_i(1)\not=I_i(1)$. Lastly, note that under ideal channels, since $z^k_n=b^k_n, \forall k,n$, we have $\tilde{I}_i(1)=\hat{I}_i(1)$.

\subsection{Asymptotic Analysis of Detection Delay}

Let $\{\tau^k_n:\tau^k_n=t^k_n-t^k_{n-1}\}$ denote the inter-arrival times of the LLR messages transmitted from the $k$-th sensor. Note that $\tau^k_n$ depends on the observations $y^k_{t^k_{n-1}+1},\ldots,y^k_{t^k_n}$, and since
$\{y^k_t\}$ are i.i.d., $\{\tau^k_n\}$ are also i.i.d. random variables. Hence, the counting process $\{N^k_t\}$ is a renewal process. Similarly the LLRs $\{\hat{\lambda}^k_n\}$ of the received signals at the FC are also i.i.d. random variables, and form a renewal-reward process.
Note from (\ref{eq:stop_time}) that the SPRT can stop in between two arrival times of sensor $k$, e.g., $t^k_n \leq \cT < t^k_{n+1}$.
The event $N^k_{\cT}=n$ occurs if and only if $t^k_n=\tau^k_1+\ldots+\tau^k_n \leq \cT$ and $t^k_{n+1}=\tau^k_1+\ldots+\tau^k_{n+1} > \cT$, so it depends on the first $(n+1)$ LLR messages. From the definition of \emph{stopping time} \cite[pp. 104]{Ross96} we conclude that $N^k_{\cT}$ is not a stopping time for the processes $\{\tau^k_n\}$ and $\{\hat{\lambda}^k_n\}$ since it depends on the ($n+1$)-th message. However, $N^k_{\cT}+1$ is a stopping time for $\{\tau^k_n\}$ and $\{\hat{\lambda}^k_n\}$ since we have $N^k_{\cT}+1=n \iff N^k_{\cT}=n-1$ which depends only on the first $n$ LLR messages. Hence, from Wald's identity \cite[pp. 105]{Ross96} we can directly write the following equalities
\begin{align}
    \label{eq:tau}
    \Exp_i\left[ \sum_{n=1}^{N^k_{\cT}+1} \tau^k_n \right] &= \Exp_i[\tau^k_1] (\Exp_i[N^k_{\cT}]+1), \\
    \label{eq:lam}
    \text{and} \ \ \ \Exp_i\left[ \sum_{n=1}^{N^k_{\cT}+1} \hat{\lambda}^k_n \right] &= \Exp_i[\hat{\lambda}^k_1] (\Exp_i[N^k_{\cT}]+1).
\end{align}
%The proof of (\ref{eq:lam}) can also be found in \cite[Lemma 3]{Fellouris11}.

We have the following theorem on the average decision delay under ideal channels.

\begin{thm} \label{thm:id1}
Consider the decentralized detection scheme given in Section \ref{sec:back}, with ideal channels between sensors and the FC. Its average decision delay under $\Hyp_i$ is given by
\be
\label{eq:id_thm1}
\Exp_i[\cT] = \frac{\hat{I}_i(\cT)}{ \hat{I}_i(1) } + \frac{\sum_{k=1}^K \hat{I}^k_i(t^k_{N^k_{\cT}+1}) - \Exp_i[\cY_k] \hat{I}^k_i(1)}{ \hat{I}_i(1) }
\ee
where $\cY_k$ is a random variable representing the time interval between the stopping time and the arrival of the first bit from the $k$-th sensor after the stopping time, i.e., $\cY_k \triangleq t^k_{N^k_{\cT}+1}-\cT$.
\end{thm}

\begin{IEEEproof}
From (\ref{eq:tau}) and (\ref{eq:lam}) we obtain
\be
\label{eq:perf_id1}
\Exp_i\left[ \sum_{n=1}^{N^k_{\cT}+1} \tau^k_n \right] = \Exp_i[\tau^k_1] \frac{\Exp_i\left[ \sum_{n=1}^{N^k_{\cT}+1} \hat{\lambda}^k_n \right] }{\Exp_i[\hat{\lambda}^k_1]} \nn
\ee
where the left-hand side equals to $\Exp_i[\cT]+\Exp_i[\cY_k]$. Note that $\Exp_i[\tau^k_1]$ is the expected stopping time of the local SPRT at the $k$-th sensor and by Wald's identity it is given by $\Exp_i[\tau^k_1]=\frac{\Exp_i[\lambda^k_1]}{\Exp_i[l^k_1]}$, provided that $\Exp_i[l^k_1]\not=0$. Hence, we have
\be
\label{eq:perf_id2}
\Exp_i[\cT]  = \frac{\Exp_i[\lambda^k_1]}{\Exp_i[\hat{\lambda}^k_1]} \frac{\Exp_i\left[ \sum_{n=1}^{N^k_{\cT}+1} \hat{\lambda}^k_n \right] }{\Exp_i[l^k_1]}-\Exp_i[\cY_k] = \frac{I^k_i(t^k_1)}{\hat{I}^k_i(t^k_1)} \frac{\hat{I}^k_i(\cT)+\hat{I}^k_i(t^k_{N^k_{\cT}+1}) }{I^k_i(1)} - \Exp_i[\cY_k] \nn
\ee
where we used the fact that $\Exp_1\left[ \sum_{n=1}^{N^k_{\cT}+1} \hat{\lambda}^k_n \right] = \Exp_1[\hat{L}^k_{\cT}]+\tilde{\Exp}_1[\hat{\lambda}^k_{N^k_{\cT}+1}] = \hat{I}^k_1(\cT)+\hat{I}^k_1(t^k_{N^k_{\cT}+1})$ and similarly $\Exp_0\left[ \sum_{n=1}^{N^k_{\cT}+1} \hat{\lambda}^k_n \right] = -\hat{I}^k_0(\cT)-\hat{I}^k_0(t^k_{N^k_{\cT}+1})$. Note that $\tilde{\Exp}_i[\cdot]$ is the expectation with respect to $\hat{\lambda}^k_{N^k_{\cT}+1}$ and $N^k_{\cT}$ under $\Hyp_i$.
By rearranging the terms and then summing over $k$ on both sides, we obtain
\be
\label{eq:perf_id3}
\Exp_i[\cT] \underbrace{\sum_{k=1}^K I^k_i(1) \frac{\hat{I}^k_i(t^k_1)}{I^k_i(t^k_1)}}_{\hat{I}_i(1)} = \hat{I}_i(\cT) + \sum_{k=1}^K \hat{I}^k_i(t^k_{N^k_{\cT}+1}) - \Exp_i[\cY_k] \underbrace{I^k_i(1) \frac{\hat{I}^k_i(t^k_1)}{I^k_i(t^k_1)}}_{\hat{I}^k_i(1)} \nn
\ee
which is equivalent to (\ref{eq:id_thm1}).
\end{IEEEproof}

The result in (\ref{eq:id_thm1}) is in fact very intuitive. Recall that $\hat{I}_i(\cT)$ is the KL information at the detection time at the FC. It naturally lacks some local information that has been accumulated at sensors, but has not been transmitted to the FC, i.e., the information gathered at sensors after their last sampling times. The numerator of the second term on the right hand side of (\ref{eq:id_thm1}) replaces such missing information by using the hypothetical KL information. Note that in (\ref{eq:id_thm1}) $\hat{I}^k_i(t^k_{N^k_{\cT}+1}) \not= \hat{I}^k_i(t^k_1)$, i.e., $\tilde{\Exp}_i[\hat{\lambda}^k_{N^k_{\cT}+1}]  \not= \Exp_i[\hat{\lambda}^k_1]$, since $N^k_{\cT}$ and $\hat{\lambda}^k_{N^k_{\cT}+1}$ are not independent.

The next result gives the asymptotic decision delay performance under ideal channels.

\begin{thm} \label{thm:id2}
As the error probability bounds tend to zero, i.e., $\alpha,\beta\to0$, the average decision delay under ideal channels given by (\ref{eq:id_thm1}) satisfies
\be
\label{eq:id_thm2}
\Exp_1[\cT] = \frac{|\log\alpha|}{ \hat{I}_1(1) } + O(1),~~\text{and}~~\Exp_0[\cT] = \frac{|\log\beta|}{ \hat{I}_0(1) } + O(1),
\ee
where $O(1)$ represents a constant term.
\end{thm}

\begin{IEEEproof}
We will prove the first equality in (\ref{eq:id_thm2}), and the proof of the second one follows similarly. Let us first prove the following lemma.
\begin{lem}
\label{lem:id}
  As $\alpha,\beta\to0$ we have the following KL information at the FC
  \be
  \label{eq:lem_id}
      \hat{I}_1(\cT)=|\log\alpha|+O(1),~~\text{and}~~\hat{I}_0(\cT)=|\log\beta|+O(1).
  \ee
\end{lem}

\begin{IEEEproof}
  We will show the first equality and the second one follows similarly. We have
  \begin{align}
  \label{eq:lem_id1}
  \hat{I}_1(\cT) =& \Pro_1(\hat{L}_{\cT} \geq A) \Exp_1[\hat{L}_{\cT}|\hat{L}_{\cT} \geq A] + \Pro_1(\hat{L}_{\cT} \leq -B) \Exp_1[\hat{L}_{\cT}|\hat{L}_{\cT} \leq -B] \nn\\
  =& (1-\beta)(A+\Exp_1[\theta_{A}]) - \beta(B+\Exp_1[\theta_{B}])
  \end{align}
where $\theta_{A}, \theta_{B}$ are overshoot and undershoot respectively given by $\theta_{A}\triangleq \hat{L}_{\cT}-A$ if $\hat{L}_{\cT}\geq A$ and $\theta_{B}\triangleq -\hat{L}_{\cT}-B$ if $\hat{L}_{\cT}\leq -B$. From \cite[Theorem 2]{Fellouris11}, we have $A \leq |\log\alpha|$ and $B \leq |\log\beta|$, so as $\alpha,\beta\to0$ (\ref{eq:lem_id1}) becomes $\hat{I}_1(\cT) = A+\Exp_1[\theta_{A}]+o(1)$.
From (\ref{eq:ideal2}) we have $|\hat{\lambda}^k_n|<\infty$ if $0<\alpha_k,\beta_k<1$. If we assume $0<\Delta<\infty$ and $|l^k_t|<\infty, \forall k,t$, then we have $0<\alpha_k,\beta_k<1$ and as a result $\hat{I}^k_i(t^k_1)=\Exp_i[\hat{\lambda}^k_1]<\infty$. Since the overshoot cannot exceed the last received LLR value, we have $\theta_{A},\theta_{B} \leq \Theta= \max_{k,n} |\hat{\lambda}^k_n| <\infty$.
Similar to Eq. (73) in \cite{Fellouris11} we can write  $\beta \geq e^{-B-\Theta}$ and $\alpha \geq e^{-A-\Theta}$ where $\Theta=O(1)$ by the above argument, or equivalently, $B\geq |\log \beta|-O(1)$ and $A\geq |\log \alpha|-O(1)$. Hence we have $A= |\log \alpha| + O(1)$ and $B = |\log \beta| + O(1)$.
\end{IEEEproof}

From the assumption of $|l^k_t|<\infty, \forall k,t$, we also have $\hat{I}_i(1)\leq I_i(1)<\infty$. Moreover, we have $\Exp_i[\cY_k] \leq \Exp_i[\tau^k_1]<\infty$ since $\Exp_i[l^k_1]\not=0$. Note that all the terms on the right-hand side of (\ref{eq:id_thm1}) except for $\hat{I}_i(\cT)$ do not depend on the global error probabilities $\alpha,\beta$, so they are $O(1)$ as $\alpha,\beta \to 0$. Finally, substituting (\ref{eq:lem_id}) into (\ref{eq:id_thm1}) we get (\ref{eq:id_thm2}).

\end{IEEEproof}

It is seen from (\ref{eq:id_thm2}) that the hypothetical KL information number, $\hat{I}_i(1)$, plays a key role in the asymptotic decision delay expression. In particular, we need to maximize $\hat{I}_i(1)$ to asymptotically minimize $\Exp_i[\cT]$. Recalling its definition
$$
\hat{I}_i(1)=\sum_{k=1}^K \frac{\hat{I}^k_i(t^k_1)}{I^k_i(t^k_1)} I^k_i(1)
$$
we see that three information numbers are required to compute it. Note that $I^k_i(1)=\Exp_i[l^k_1]$ and $I^k_i(t^k_1)=\Exp_i[\lambda^k_1]$, which is given in (\ref{eq:inf_num_sen}) below, are computed based on local observations at sensors, thus do not depend on the channels between sensors and the FC. Specifically, we have
\begin{align}
\label{eq:inf_num_sen}
  \begin{split}
    I^k_1(t^k_1) &= (1-\beta_k) (\Delta+\Exp_1[\bar{\theta}^k_n]) - \beta_k (\Delta+\Exp_1[\underline{\theta}^k_n]), \\
    \text{and} \ \ \ I^k_0(t^k_1) &= \alpha_k (\Delta+\Exp_0[\bar{\theta}^k_n]) - (1-\alpha_k) (\Delta+\Exp_0[\underline{\theta}^k_n])
  \end{split}
\end{align}
where $\bar{\theta}^k_n$ and $\underline{\theta}^k_n$ are local over(under)shoots given by $\bar{\theta}^k_n\triangleq \lambda^k_n-\Delta$ if $\lambda^k_n \geq \Delta$ and $\underline{\theta}^k_n \triangleq -\lambda^k_n-\Delta$ if $\lambda^k_n \leq -\Delta$. Due to having $|l^k_t|<\infty, \forall k,t$ we have $\bar{\theta}^k_n, \underline{\theta}^k_n < \infty, \forall k,n$.
%One needs to first compute the local error probabilities $\alpha_k$ and $\beta_k$, and the expected local overshoots $\Exp_i[\bar{\theta}^k_n]$ and $\Exp_i[\underline{\theta}^k_n]$ to obtain the information number $I^k_i(t^k_1)$.

On the other hand, $\hat{I}^k_i(t^k_1)$ represents the information received in an LLR message by the FC, so it heavily depends on the channel type. In the ideal channel case, from (\ref{eq:ideal2}) it is given by
\begin{align}
\label{eq:inf_num_fc}
  \begin{split}
    \hat{I}^k_1(t^k_1)&= (1-\beta_k) \log\frac{1-\beta_k}{\alpha_k} + \beta_k \log\frac{\beta_k}{1-\alpha_k}, \\
    \text{and} \ \ \ \hat{I}^k_0(t^k_1)&= \alpha_k \log\frac{1-\beta_k}{\alpha_k} + (1-\alpha_k) \log\frac{\beta_k}{1-\alpha_k}.
  \end{split}
\end{align}
Since $\hat{I}^k_i(t^k_1)$ is the only channel-dependent term in the asymptotic decision delay expression, in the next section we will obtain its expression for each noisy channel type considered in Section \ref{sec:Alg}.

\section{Performance Analysis for Noisy Channels}
\label{sec:Per_noi}

In all noisy channel types that we consider in this paper, we assume that channel parameters are either constants or i.i.d. random variables across time. In other words, $\epsilon_k,h_k$ are constant for all $k$ (see Section \ref{sec:bec_alg}, \ref{sec:bsc_alg}, \ref{sec:awgn_alg}), and $\{h^k_n\}_n,\{w^k_n\}_n$ are i.i.d. for all $k$ (see Section \ref{sec:awgn_alg}, \ref{sec:ray_alg}, \ref{sec:ric_alg}). Thus, in all noisy channel cases discussed in Section \ref{sec:Alg} the inter-arrival times of the LLR messages $\{\tilde{\tau}^k_n\}$, and the LLRs of the received signals $\{\tilde{\lambda}^k_n\}$ are i.i.d. across time as in the ideal channel case. Accordingly the average decision delay in these noisy channels has the same expression as (\ref{eq:id_thm1}), as given by the following proposition. The proof is similar to that of Theorem \ref{thm:id1}.

\begin{pro}
\label{pro:noi1}
Under each type of noisy channel discussed in Section \ref{sec:Alg}, the average decision delay is given by
\be
\label{eq:perf_noi1}
\Exp_i[\tilde{\cT}] = \frac{\tilde{I}_i(\tilde{\cT})}{ \tilde{I}_i(1) } + \frac{\sum_{k=1}^K  \tilde{I}^k_i(t^k_{N^k_{\cT}+1}) - \Exp_i[\tilde{\cY}_k] \tilde{I}^k_i(1) }{ \tilde{I}_i(1) }
\ee
where $\tilde{\cY}_k \triangleq t^k_{N^k_{\tilde{\cT}}+1}-\tilde{\cT}$.
\end{pro}

The asymptotic performances under noisy channels can also be analyzed analogously to the ideal channel case.

\begin{pro}
\label{pro:noi2}
As $\alpha, \beta \to 0$, the average decision delay under noisy channels given by (\ref{eq:perf_noi1}) satisfies
\be
\label{eq:delay_noi}
\Exp_1[\tilde{\cT}] = \frac{|\log\alpha|}{ \tilde{I}_1(1) } + O(1),~~\text{and}~~\Exp_0[\tilde{\cT}] = \frac{|\log\beta|}{ \tilde{I}_0(1) } + O(1).
\ee
\end{pro}

\begin{IEEEproof}
Note that in the noisy channel cases the FC, as discussed in Section \ref{sec:Alg}, computes the LLR, $\tilde{\lambda}^k_n$, of the signal it receives, and then performs SPRT using the LLR sum $\tilde{L}_t$. Hence, analogous to Lemma \ref{lem:id} we can show that $\tilde{I}_1(\tilde{\cT})=|\log\alpha|+O(1)$ and $\tilde{I}_0(\tilde{\cT})=|\log\beta|+O(1)$ as $\alpha,\beta\to0$.
Note also that due to channel uncertainties $|\tilde{\lambda}^k_n|\leq|\hat{\lambda}^k_n|$, so we have $\tilde{I}^k_i(t^k_1)\leq\hat{I}^k_i(t^k_1) < \infty$ and $\tilde{I}_i(1)\leq\hat{I}_i(1) < \infty$. We also have $\Exp_i[\tilde{\cY}_k] \leq \Exp_i[\tilde{\tau}^k_1] < \infty$ as in the ideal channel case. Substituting these asymptotic values in (\ref{eq:perf_noi1}) we get (\ref{eq:delay_noi}).
\end{IEEEproof}

Recall that $\tilde{I}_i(1)=\sum_{k=1}^K \frac{\tilde{I}^k_i(t^k_1)}{I^k_i(t^k_1)} I^k_i(1)$ in (\ref{eq:delay_noi}) where $I^k_i(1)$ and $I^k_i(t^k_1)$ are independent of the channel type, i.e., they are same as in the ideal channel case. In the subsequent subsections, we will compute $\tilde{I}^k_i(t^k_1)$ for each noisy channel type. We will also consider the choices of the signaling levels $a, b$ in (\ref{eq:tx_sig}) that maximize $\tilde{I}^k_i(t^k_1)$.

\subsection{BEC}

Under BEC, from (\ref{eq:bec1}) we can write the LLR of the received bits at the FC as
    \bea
    \tilde{\lambda}^k_n &=& \left\{ \begin{array}{ll}
            \hat{\lambda}^k_n, & \mbox{with probability $1-\epsilon_k$}, \\
            0, & \mbox{with probability $\epsilon_k$}.
            \end{array} \right.
    \eea
Hence we have
    \be
    \label{eq:bec_perf1}
    \tilde{I}^k_i(t^k_1) = \Exp_i[\tilde{\lambda}^k_1] = (1-\epsilon_k) \hat{I}^k_i(t^k_1)
    \ee
 where $\hat{I}^k_i(t^k_1)$ is given in (\ref{eq:inf_num_fc}). As can be seen in (\ref{eq:bec_perf1}) the performance degradation under BEC is only determined by the channel parameters $\epsilon_k$. In general, from (\ref{eq:id_thm2}), (\ref{eq:delay_noi}) and (\ref{eq:bec_perf1}) this asymptotic performance loss can be quantified as
 %\be
 %\label{eq:bec_perf2}
$ \frac{1}{1-\min_k \epsilon_k} \leq \frac{\Exp_i[\tilde{\cT}]}{\Exp_i[\cT]}  \leq  \frac{1}{1-\max_k \epsilon_k}$.
% \ee
Specifically, if $\epsilon_k=\epsilon, \forall k$, then we have $\frac{\Exp_i[\tilde{\cT}]}{\Exp_i[\cT]}  = \frac{1}{1-\epsilon}$ as $\alpha, \beta \to 0$.

\subsection{BSC}
\label{sec:bsc_per}

Recall from (\ref{eq:bsc2}) and (\ref{eq:bsc3}) that under BSC local error probabilities $\alpha_k,\beta_k$ undergo a linear transformation to yield the effective local error probabilities $\hat{\alpha}_k,\hat{\beta}_k$ at the FC. Therefore, using (\ref{eq:bsc2}) and (\ref{eq:bsc3}), similar to (\ref{eq:inf_num_fc}), $\tilde{I}^k_i(t^k_1)$ is written as follows
\begin{align}
\label{eq:bsc_perf1}
  \begin{split}
    \tilde{I}^k_1(t^k_1)&= (1-\hat{\beta}_k) \log\frac{1-\hat{\beta}_k}{\hat{\alpha}_k} + \hat{\beta}_k \log\frac{\hat{\beta}_k}{1-\hat{\alpha}_k},\\
    \text{and} \ \ \ \tilde{I}^k_0(t^k_1)&= \hat{\alpha}_k \log\frac{1-\hat{\beta}_k}{\hat{\alpha}_k} + (1-\hat{\alpha}_k) \log\frac{\hat{\beta}_k}{1-\hat{\alpha}_k}
  \end{split}
\end{align}
where $\hat{\alpha}_k=(1-2\epsilon_k)\alpha_k+\epsilon_k$ and $\hat{\beta}_k=(1-2\epsilon_k)\beta_k+\epsilon_k$. Notice that the performance loss in this case also depends only on the channel parameter $\epsilon_k$.

\begin{figure}[t]
\centering
\includegraphics[scale=0.4]{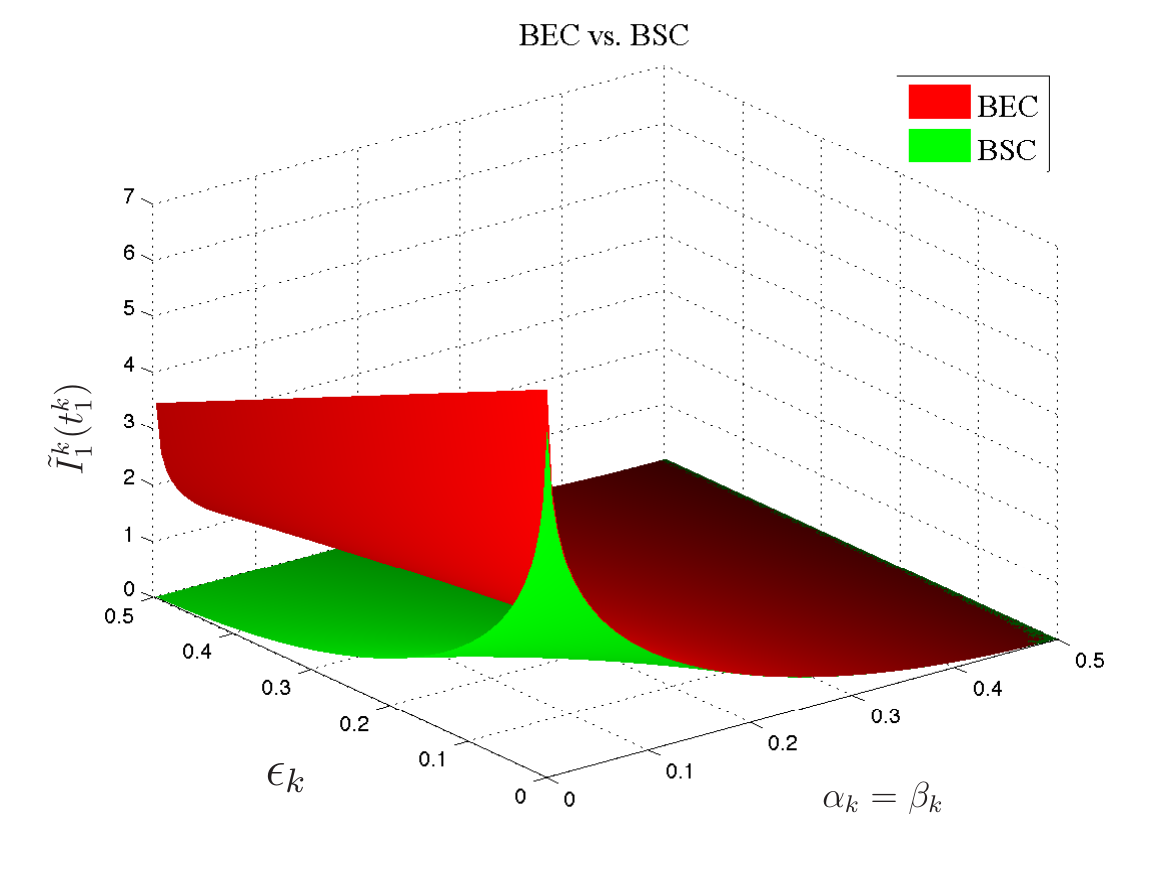}
\caption{The KL information, $\tilde{I}^k_1(t^k_1)$, under BEC and BSC, as a function of the local error probabilities $\alpha_k=\beta_k$ and the channel error probability $\epsilon_k$.}
\label{fig:BecBsc}
\end{figure}
In Fig. 2 we plot $\tilde{I}^k_1(t^k_1)$ as a function of $\alpha_k=\beta_k$ and $\epsilon_k$, for both BEC and BSC. It is seen that the KL information of BEC is higher than that of BSC, implying that the asymptotic average decision delay is lower for BEC, as anticipated in Section \ref{sec:bsc_alg}.

\subsection{AWGN}
\label{sec:awgn_per}

In this and the following sections, we will drop the sensor index $k$ of $\sigma_{h,k}^2$ and $\sigma_k^2$ for simplicity.
In the AWGN case, it follows from Section \ref{sec:awgn_alg} that if the transmitted signal is $a$, i.e., $x^k_n=a$, then $c^k_n=u, d^k_n=v_a$; and
if $x^k_n=b$, then $c^k_n=v_b, d^k_n=u$ where $u \triangleq \frac{| w^k_n |^2}{\sigma^2}, v_a \triangleq \frac{| w^k_n+(a-b)h_k |^2}{\sigma^2}, v_b \triangleq \frac{| w^k_n+(b-a)h_k |^2}{\sigma^2}$. Accordingly, from (\ref{eq:LLR_awgn}) we write the KL information as
\begin{align}
\label{eq:awgn_perf1}
\tilde{I}^k_1(t^k_1) =& \bar{\Exp}_1[\tilde{\lambda}^k_1] = (1-\beta_k) \Exp\left[ \log \frac { (1-\beta_k) e^{-u} + \beta_k e^{-v_a} } { \alpha_k e^{-u} + (1-\alpha_k) e^{-v_a} }\right] + \beta_k \Exp\left[ \log \frac { (1-\beta_k) e^{-v_b} + \beta_k e^{-u} } { \alpha_k e^{-v_b} + (1-\alpha_k) e^{-u} } \right] \nn\\
=& \underbrace{(1-\beta_k) \log\frac{1-\beta_k}{\alpha_k} + \beta_k \log\frac{\beta_k}{1-\alpha_k} }_{\hat{I}^k_1(t^k_1)} + \nn\\
& \beta_k  \underbrace{ \Big( \frac{1-\beta_k}{\beta_k} \overbrace{\Exp\left[ \log \frac { 1 + \frac{\beta_k}{1-\beta_k} e^{u-v_a} } { 1 + \frac{1-\alpha_k}{\alpha_k} e^{u-v_a} }\right]}^{\mathcal{E}_1} + \overbrace{ \Exp\left[ \log \frac { 1+\frac{1-\beta_k}{\beta_k} e^{u-v_b}} { 1+\frac{\alpha_k}{1-\alpha_k} e^{u-v_b} } \right]}^{\mathcal{E}_2} \Big) }_{\cC^k_1},
\end{align}
where $\Exp[\cdot]$ denotes the expectation with respect to the channel noise $w^k_n$ only, and $\bar{\Exp}_1[\cdot]$ denotes the expectation with respect to both $x^k_n$ and $w^k_n$ under $\Hyp_1$. Since $w^k_n$ is independent of $x^k_n$ under both $\Hyp_0$ and $\Hyp_1$, we used the identity $\bar{\Exp}_1[\cdot]=\Exp[\Exp_1[\cdot]]$ in (\ref{eq:awgn_perf1}).

Note from (\ref{eq:awgn_perf1}) that we have $\tilde{I}^k_1(t^k_1)=\hat{I}^k_1(t^k_1)+\beta_k \cC^k_1$ and $\tilde{I}^k_0(t^k_1)=\hat{I}^k_0(t^k_1)+\alpha_k \cC^k_0$. Similar to $\cC^k_1$ we have $\cC^k_0 \triangleq -\mathcal{E}_1 - \frac{1-\alpha_k}{\alpha_k} \mathcal{E}_2$. Since we know $\tilde{I}^k_i(t^k_1)\leq\hat{I}^k_i(t^k_1)$, the extra terms,  $\cC^k_1, \cC^k_0 \leq 0$ are penalty terms that correspond to the information loss due to the channel noise. Our focus will be on this term as we want to optimize the performance under AWGN channels by choosing the transmission signal levels $a$ and $b$ that maximize $\cC^k_i$.

Let us first consider the random variables $\zeta_a \triangleq u-v_a$ and $\zeta_b \triangleq u-v_b$ which are the arguments of the exponential functions in $\mathcal{E}_1$ and $\mathcal{E}_2$ in (\ref{eq:awgn_perf1}) . From the definitions of $u$ and $v_a$, we write
$\zeta_a= \frac{|w^k_n|^2}{\sigma^2} - \frac{|w^k_n+(a-b)h_k|^2}{\sigma^2} = - \frac{|a-b|^2|h_k|^2}{\sigma^2} - \frac{2}{{\sigma^2}} \gamma$
where $\gamma \triangleq \Re\{(w^k_n)^* (a-b)h_k\}$ and $\Re\{\cdot\}$ denotes the real part of a complex number. Similarly we have $\zeta_b= -\frac{|a-b|^2|h_k|^2}{\sigma^2} + \frac{2}{{\sigma^2}} \gamma$. Note that $\gamma \sim \cN(0,\frac{|a-b|^2 |h_k|^2 \sigma^2}{2})$ since $w^k_n \sim \cN_c(0,\sigma^2)$. If we define $\nu \triangleq \frac{\sqrt{2}}{|a-b| |h_k| \sigma} \gamma$, then we have $\nu \sim \cN(0,1)$. Upon defining  $s \triangleq \frac{|a-b| |h_k|}{\sigma}$ we can then write $\zeta_a$ and $\zeta_b$ as
\be
\zeta_a = -s^2 - \sqrt{2}~s\nu \ \ \ \ \text{and} \ \ \ \ \zeta_b = -s^2 + \sqrt{2}~s\nu. \nn
\ee
If we define $F \triangleq \frac{1-\alpha_k}{\alpha_k}$ and $G \triangleq \frac{1-\beta_k}{\beta_k}$, then we have
\begin{align}
\label{eq:awgn_perf2}
\begin{split}
\cC^k_0 =& \Exp\left[ \log \frac{1+F^{-1}e^{\zeta_b}}{1+Ge^{\zeta_b}} \right] + F^{-1} \Exp\left[ \log \frac{1+Fe^{\zeta_a}}{1+G^{-1}e^{\zeta_a}} \right] \\
\cC^k_1 =& \Exp\left[ \log \frac{1+Ge^{\zeta_b}}{1+F^{-1}e^{\zeta_b}} \right] + G \Exp\left[ \log \frac{1+G^{-1}e^{\zeta_a}}{1+Fe^{\zeta_a}} \right].
\end{split}
\end{align}

Note from (\ref{eq:sigmodel_awgn}) that the received signal, $z^k_n$, will have the same variance, but different means, $a h_k$ and $b h_k$, if $x^k_n=a$ and $x^k_n=b$ are transmitted respectively.
Hence, we expect that the detection performance under AWGN channels will improve if the difference between the transmission levels, $|a-b|$, increases. Toward that end the following result gives a sufficient condition under which the penalty term $\cC^k_i$ increases with $s$, and hence with $|a-b|$. The proof is given in the Appendix.

\begin{lem}
\label{lem:aw}
$\cC^k_i$ is an increasing function of $s$, $i=0,1$, if $F^2 \geq G$ and $G^2 \geq F$.
\end{lem}

\begin{figure}[t]
\centering
\includegraphics[scale=0.4]{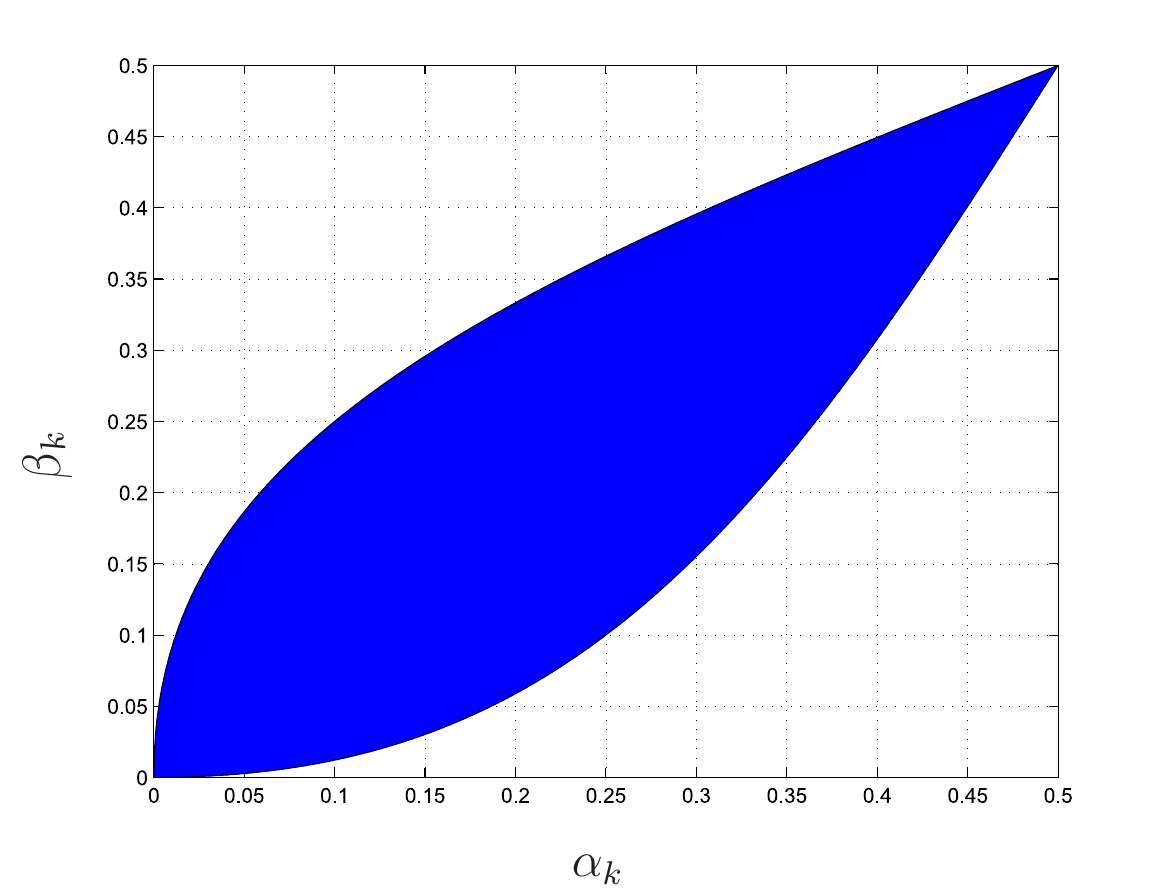}
\caption{The region of $(\alpha_k, \beta_k)$ specified by Lemma \ref{lem:aw}. }
\label{fig:awgn}
\end{figure}

Lemma \ref{lem:aw} indicates that for $\alpha_k,\beta_k$ values inside the region shown in Fig. \ref{fig:awgn}, $\cC^k_i$ is increasing in $|a-b|$.
Note that $\alpha_k,\beta_k$ are local error probabilities which are directly related to the local threshold $\Delta$. Therefore, even if the hypotheses $\Hyp_0$ and $\Hyp_1$ are non-symmetric, we can ensure that we will have $\alpha_k,\beta_k$ inside the region in Fig. \ref{fig:awgn} by employing different local thresholds, $-\underline{\Delta}_k$ and $\bar{\Delta}_k$, in (\ref{eq:LSsamptime}).
In fact, even for $\alpha_k,\beta_k$ values outside the region in Fig. \ref{fig:awgn} numerical results show that $\cC^k_i$ is increasing in $s$.

Hence, maximizing $\cC^k_i$ is equivalent to maximizing $|a-b|$. If we consider a constraint on the maximum allowed transmission power at sensors, i.e., $\max(|a|^2,|b|^2) \leq P^2$, then the antipodal signaling is optimum, i.e., $|a|=|b|=P$ and $a=-b$.

\subsection{Rayleigh Fading}
\label{sec:ray_per}

It follows from Section \ref{sec:ray_alg} that $c^k_n=u_a,~d^k_n=\frac{\sigma_a^2}{\sigma_b^2}u_a$ when $x^k_n=a$; and $c^k_n=\frac{\sigma_b^2}{\sigma_a^2}u_b,~d^k_n=u_b$ when $x^k_n=b$ where $u_a \triangleq \frac{|a h^k_n+w^k_n|^2}{\sigma_a^2}$, $u_b \triangleq \frac{|b h^k_n+w^k_n|^2}{\sigma_b^2}$, and $\sigma_a^2 = |a|^2 \sigma_h^2 + \sigma^2$,  $\sigma_b^2 = |b|^2 \sigma_h^2 + \sigma^2$ as defined in Section \ref{sec:ray_alg}. Define further $\rho \triangleq \frac{\sigma_a^2}{\sigma_b^2}$. Hence, using (\ref{eq:LLR_ray}) we write the KL information as
\begin{align}
\label{eq:ray_perf1}
\tilde{I}^k_1(t^k_1) =& (1-\beta_k) \Exp\left[ \log \frac { \frac{1-\beta_k}{\sigma_a^2} e^{-u_a} + \frac{\beta_k}{\sigma_b^2} e^{-\rho u_a} } { \frac{\alpha_k}{\sigma_a^2} e^{-u_a} + \frac{1-\alpha_k}{\sigma_b^2} e^{-\rho u_a} }\right] + \beta_k \Exp\left[ \log \frac { \frac{1-\beta_k}{\sigma_a^2} e^{-\rho^{-1} u_b} + \frac{\beta_k}{\sigma_b^2} e^{-u_b} } { \frac{\alpha_k}{\sigma_a^2} e^{-\rho^{-1} u_b} + \frac{1-\alpha_k}{\sigma_b^2} e^{-u_b} } \right] \nn\\
=& \underbrace{(1-\beta_k) \log\frac{1-\beta_k}{\alpha_k} + \beta_k \log\frac{\beta_k}{1-\alpha_k} }_{\hat{I}^k_1(t^k_1)} +  \beta_k \underbrace{ \left( \Exp\left[ \log \frac{1+G \rho^{-1} e^{\zeta_b}}{1+F^{-1} \rho^{-1} e^{\zeta_b}} \right] + G \Exp\left[ \log \frac{1+G^{-1} \rho e^{\zeta_a}}{1+F \rho e^{\zeta_a}} \right] \right)}_{\cC^k_1}
\end{align}
where $\zeta_a \triangleq u_a(1-\rho)$ and $\zeta_b \triangleq u_b(1-\rho^{-1})$. Note that when $|a|=|b|$ which corresponds to the optimal signaling in the AWGN case, we have $\rho=1$, $\zeta_a=\zeta_b=0$ and therefore $\tilde{I}^k_1(t^k_1)=0$ in (\ref{eq:ray_perf1}). This result is quite intuitive since in the Rayleigh fading case the received signals differ only in their variances. Note that $u_a$ and $u_b$ are chi-squared random variables with $2$ degrees of freedom, i.e., $u_a, u_b \sim \chi^2_2$, thus we can write the penalty term $\cC^k_i$ as
\begin{align}
\label{eq:ray_perf2}
\begin{split}
\cC^k_0 =& \int_0^{\infty} \left( \log  \frac{1+F^{-1} \rho^{-1} e^{u(1-\rho^{-1})}}{1+G \rho^{-1} e^{u(1-\rho^{-1})}} + F^{-1} \log     \frac{1+F \rho e^{u(1-\rho)}}{1+G^{-1} \rho e^{u(1-\rho)}}    \right)    \frac{e^{-u}}{2} \text{d} u, \\
\text{and} \ \ \ \cC^k_1 =& \int_0^{\infty} \left( \log   \frac{1+G \rho^{-1} e^{u(1-\rho^{-1})}}{1+F^{-1} \rho^{-1} e^{u(1-\rho^{-1})}} + G \log   \frac{1+G^{-1} \rho e^{u(1-\rho)}}{1+F \rho e^{u(1-\rho)}}    \right)    \frac{e^{-u}}{2} \text{d} u.
\end{split}
\end{align}

\begin{figure}[t]
\centering
\includegraphics[scale=0.65]{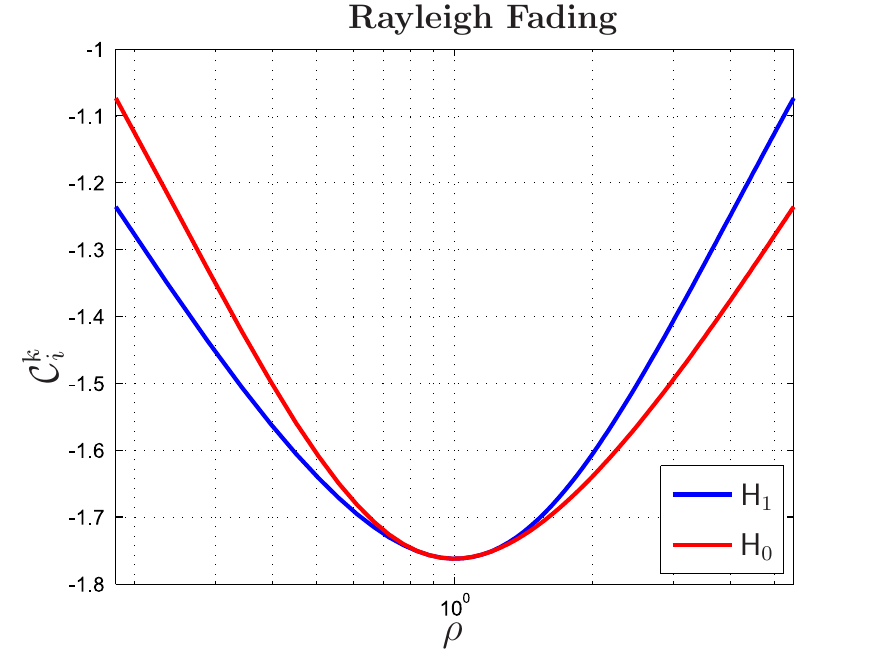}
\caption{The penalty term $\cC^k_i$ for Rayleigh fading channels as a function of $\rho$.}
\label{fig:ray_rho}
\end{figure}

Note that given local error probabilities $\alpha_k, \beta_k$ the integrals in (\ref{eq:ray_perf2}) is a function of $\rho$ only. However, maximizing $\cC^k_i$ in (\ref{eq:ray_perf2}) with respect to $\rho$ seems analytically intractable.
As can be seen in Section \ref{sec:ray_alg}, the received signals at the FC will have zero mean and the variances $\sigma_a^2$ and $\sigma_b^2$ when $x^k_n=a$ and $x^k_n=b$ respectively. Therefore, in this case intuitively we should increase the difference between the two variances, i.e., $\left| |a|^2-|b|^2 \right|$.
Consider the following constraints: $\max(|a|^2,|b|^2) \leq P^2$ and $\min(|a|^2,|b|^2) \geq Q^2$, where the first one is the peak power constraint as before, and the second is to ensure reliable detection of an incoming signal by the FC. We conjecture that the optimum signaling scheme in this case that maximizes $\cC_i^k$ corresponds to $|a|=P, |b|=Q$ or $|a|=Q, |b|=P$.

To numerically illustrate the behavior of $\cC_i^k$ as a function of $\rho$, we set $\alpha_k=\beta_k=0.1$, $\sigma_h^2=\sigma^2=1$, $P^2=10$, $Q^2=1$, and plot $\cC_i^k$ in Fig. \ref{fig:ray_rho}.
It is seen that $\cC^k_i$ has its global minimum when $\rho=1$, which corresponds to the case $|a|=|b|$ as expected. Moreover, $\cC^k_i$, validating our conjecture, monotonically grows as $\rho$ tends to its minimum and maximum values corresponding to the cases $|a|=Q, |b|=P$ and $|a|=P, |b|=Q$ respectively.

Note that in Fig. \ref{fig:ray_rho}, the curves for $\Hyp_0$ and $\Hyp_1$ are mirrored versions of each other around $\rho=1$ since we have $\alpha_k=\beta_k$ in the example. From (\ref{eq:ray_perf2})  we can say that the symmetry between $\Hyp_0$ and $\Hyp_1$ around $\rho=1$ will exist whenever $F=G$, i.e., $\alpha_k=\beta_k$.

\subsection{Rician Fading}
\label{sec:ric_per}

In the Rician fading case, upon defining $\tilde{h}^k_n \triangleq h^k_n-\mu_k$
from Section \ref{sec:ric_alg} we have $c^k_n = \frac{| a\tilde{h}^k_n+w^k_n |^2}{\sigma_a^2}$, $d^k_n = \frac{| a\tilde{h}^k_n+w^k_n + (a-b)\mu_k |^2}{\sigma_b^2}$ when $x^k_n=a$; and $c^k_n = \frac{| b\tilde{h}^k_n+w^k_n + (b-a)\mu_k |^2}{\sigma_a^2}$, $d^k_n = \frac{| b\tilde{h}^k_n+w^k_n|^2}{\sigma_b^2}$ when $x^k_n=b$. We will drop the subscript $k$ in $\mu_k$ for convenience.
We further define $\tilde{z}_a \triangleq a\tilde{h}^k_n+w^k_n$ and $\tilde{z}_b \triangleq b\tilde{h}^k_n+w^k_n$ that are zero-mean Gaussian variables with variances $\sigma_a^2$ and $\sigma_b^2$, respectively. Then from Section \ref{sec:ric_alg} similar to (\ref{eq:ray_perf1}) we write the KL information as
\begin{align}
\label{eq:ric_perf1}
\tilde{I}^k_1(t^k_1) =& \hat{I}^k_1(t^k_1) + \beta_k \underbrace{ \left( \Exp\left[ \log \frac{1+G \rho^{-1} e^{\zeta_b}}{1+F^{-1} \rho^{-1} e^{\zeta_b}} \right] + G \Exp\left[ \log \frac{1+G^{-1} \rho e^{\zeta_a}}{1+F \rho e^{\zeta_a}} \right] \right)}_{\cC^k_1}
\end{align}
where $\zeta_a \triangleq -\left( \frac{|\tilde{z}_a+(a-b)\mu|^2}{\sigma_b^2}-\frac{|\tilde{z}_a|^2}{\sigma_a^2} \right)$ and $\zeta_b \triangleq -\left( \frac{|\tilde{z}_b+(b-a)\mu|^2}{\sigma_a^2}-\frac{|\tilde{z}_b|^2}{\sigma_b^2} \right)$. Now we will analyze the exponents $\zeta_a$ and $\zeta_b$.

\subsubsection{Case 1: $|a|\not=|b|$}

\begin{figure}[t]
\centering
\includegraphics[scale=0.4]{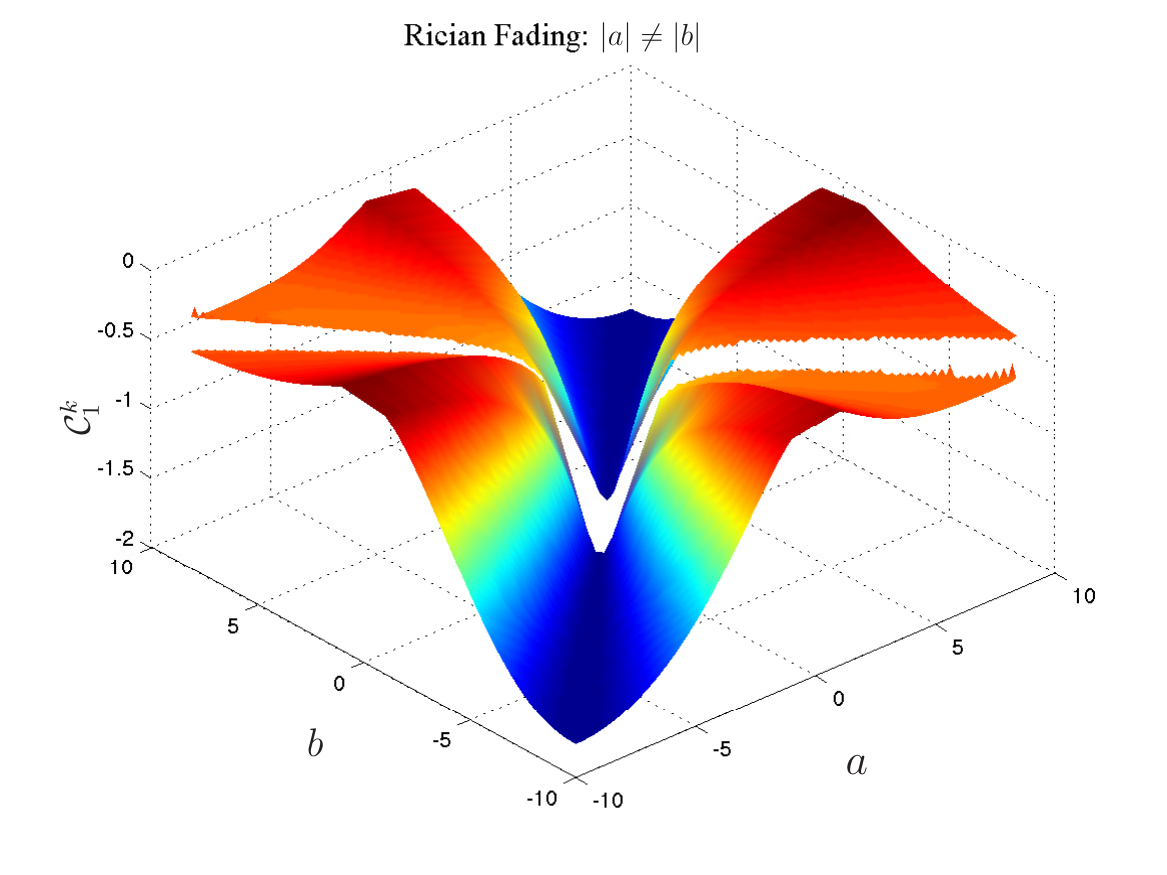} ~~
\includegraphics[scale=0.4]{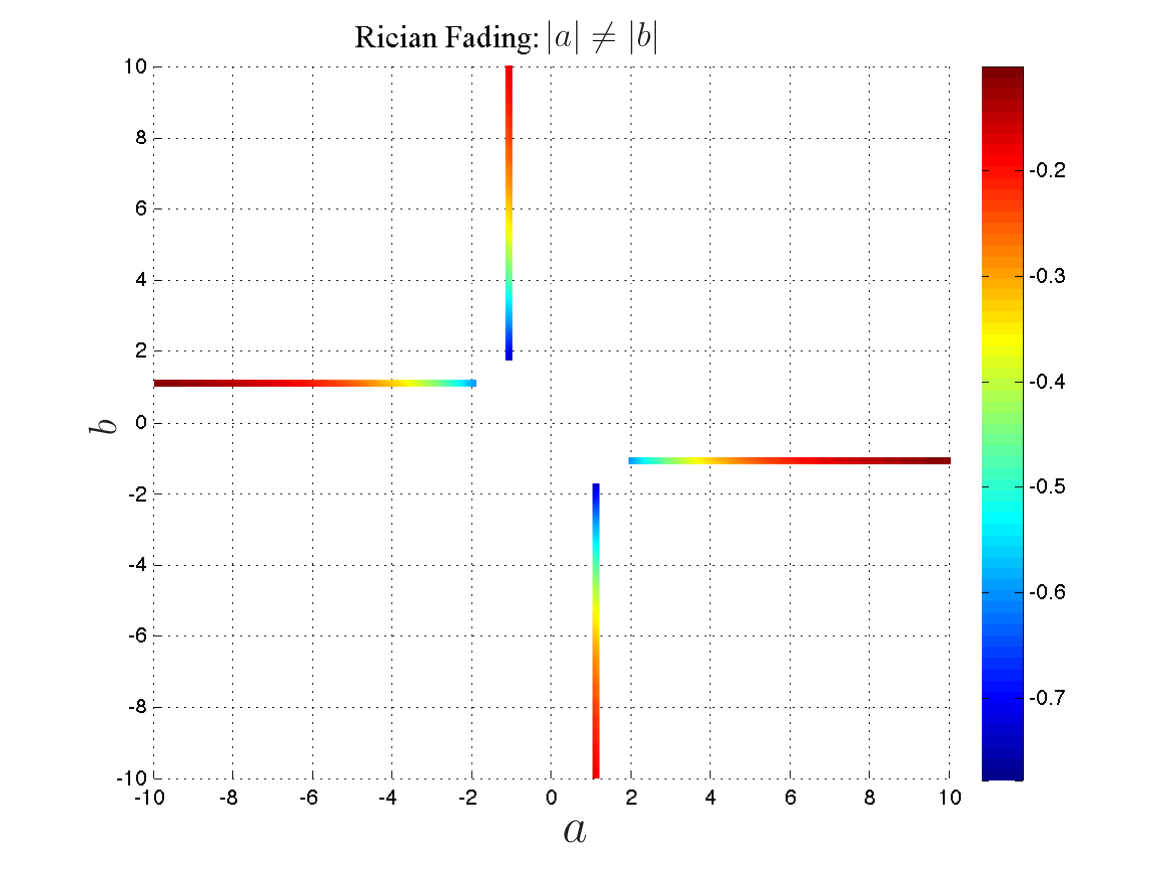}
\caption{(a) The penalty term under $\Hyp_1$, i.e., $\cC^k_1$, for Rician channels with $|a| \neq |b|$, as a function of the transmission levels $a$ and $b$. (b) The maximum contour is shown to exhibit the locus of the optimum signaling levels. $\cC^k_1$ is color coded according to the color bar given next to the figure.}
\label{fig:ric_nonsym}
\end{figure}

\ \ \ For $\kappa \triangleq \frac{1}{\sigma_b^2}-\frac{1}{\sigma_a^2} >0$, i.e., $|a|>|b|$, we can write $\zeta_a$ as
\begin{align}
\label{eq:ric_perf2}
\zeta_a =& - \Big[  \kappa |\tilde{z}_a|^2 + \frac{2 \Re\{[\tilde{z}_a^* (a-b)\mu]\}}{\sigma_b^2} + \frac{|a-b|^2|\mu|^2}{\sigma_b^2} \Big] \\
=& -\left( |\sqrt{\kappa} \tilde{z}_a|^2 + \frac{2 \sqrt{\kappa} \Re\{\tilde{z}_a^* (a-b)\mu\}}{\sigma_b^2 \sqrt{\kappa}} + \frac{|a-b|^2|\mu|^2}{\sigma_b^4 \kappa} + \frac{|a-b|^2|\mu|^2}{\sigma_b^2} - \frac{|a-b|^2|\mu|^2}{\sigma_b^4 \kappa} \right) \nn\\
=& -\Big(  \left| \sqrt{\kappa} \tilde{z}_a + \frac{(a-b)\mu}{\sigma_b^2 \sqrt{\kappa}} \right|^2 + \frac{|a-b|^2|\mu|^2 (\sigma_b^2 \kappa-1)}{\sigma_b^4 \kappa}  \Big) \nn\\
\label{eq:ric_per3}
=& -\Big(  \sigma_a^2 \kappa \underbrace{ \left| \frac{ \tilde{z}_a}{ \sigma_a} + \frac{(a-b)\mu}{\sigma_b^2 \sigma_a \kappa} \right|^2 }_{\triangleq u_a} + \frac{|a-b|^2|\mu|^2}{(|b|^2-|a|^2)\sigma_h^2}  \Big) \\
=& u_a (1-\rho) + \frac{|a-b|^2|\mu|^2}{(|a|^2-|b|^2)\sigma_h^2}
\label{eq:ric_perf4}
\end{align}
where we used $\sigma_b^2 \kappa-1=-\rho^{-1}$, $\sigma_b^4 \kappa=\rho^{-1}(\sigma_a^2-\sigma_b^2)$ while writing (\ref{eq:ric_per3}), and $\sigma_a^2 \kappa=\rho-1$ while writing (\ref{eq:ric_perf4}). Note that $u_a$ is a noncentral chi-squared random variable with two degrees of freedom and the noncentrality parameter $\lambda_a \triangleq \frac{|a-b|^2 |\mu|^2 \sigma_a^2 }{(|a|^2-|b|^2)^2 \sigma_h^4}$. Using $\sqrt{-\kappa}$ instead of $\sqrt{\kappa}$ it can be easily shown that (\ref{eq:ric_perf4}) holds for $\kappa<0$. Similarly one can obtain
\be
\zeta_b = u_b (1-\rho^{-1}) + \frac{|b-a|^2|\mu|^2}{(|b|^2-|a|^2)\sigma_h^2} \nn
\ee
for $\kappa \not= 0$, i.e., $|a| \not= |b|$, where $u_b \triangleq \left| \frac{ \tilde{z}_b}{ \sigma_b} + \frac{(a-b)\mu}{\sigma_a^2 \sigma_b \kappa} \right|^2$ and $u_b \sim \chi^2_2(\lambda_b)$ with $\lambda_b \triangleq \frac{|a-b|^2 |\mu|^2 \sigma_b^2 }{(|a|^2-|b|^2)^2 \sigma_h^4}$. Accordingly, for the non-symmetric case where $|a|\not= |b|$ from (\ref{eq:ric_perf1}) we can write $\cC^k_1$ as
\begin{align}
\label{eq:ric_nonsym}
\cC^k_1 =& \underbrace{ \int_0^{\infty} \log \left[ \frac{1+G \rho^{-1} e^{u (1-\rho^{-1}) + \frac{|b-a|^2|\mu|^2}{(|b|^2-|a|^2)\sigma_h^2} }  }  { 1+F^{-1} \rho^{-1} e^{u (1-\rho^{-1}) + \frac{|b-a|^2|\mu|^2}{(|b|^2-|a|^2)\sigma_h^2} }  }   \right]    \frac{e^{-\frac{u + \lambda_b  } {2}}}  {2}  I_0\left(  \sqrt{\lambda_b u }  \right)    \text{d} u }_{\mathcal{I}_1} +  \nn\\
& G \underbrace{ \int_0^{\infty} \log \left[ \frac{1+G^{-1} \rho e^{u (1-\rho) + \frac{|a-b|^2|\mu|^2}{(|a|^2-|b|^2)\sigma_h^2} }  }    { 1+F \rho e^{u (1-\rho) + \frac{|a-b|^2|\mu|^2}{(|a|^2-|b|^2)\sigma_h^2} }  }   \right]     \frac{e^{-\frac{u +\lambda_a  } {2}}}  {2}  I_0\left(  \sqrt{\lambda_a u }  \right)    \text{d} u }_{\mathcal{I}_2}.
\end{align}
Similarly, we have $\cC^k_0= -\mathcal{I}_1- F^{-1} \mathcal{I}_2$.

The expression in (\ref{eq:ric_nonsym}) resembles the one in (\ref{eq:ray_perf2}) for the Rayleigh fading case. And maximizing (\ref{eq:ric_nonsym}) analytically with respect to $a$ and $b$ seems even more intractable. Recall that in the Rayleigh fading case, the optimum signaling scheme was an OOK-like non-symmetric constellation, i.e., $|a|=P, |b|=Q$ or $|a|=Q, |b|=P$. Considering the same power constraints we conjecture that the same signaling scheme, that maximizes the difference between the variances $\sigma_a^2$ and $\sigma_b^2$, is optimum in this non-symmetric case.

We provide a numerical example to illustrate the behavior of $\cC^k_i$ as a function of $a$ and $b$. Using the same values for $\alpha_k, \beta_k, \sigma_h^2, \sigma^2, P^2, Q^2$ as in the Rayleigh fading case, and setting $\mu=1+j$ we plot $\cC^k_1$ in Fig. \ref{fig:ric_nonsym}(a).
The maximum contour of the three-dimensional surface in Fig. \ref{fig:ric_nonsym}(a), which corresponds to the potential optimum signaling level pairs, is clearly shown in Fig. \ref{fig:ric_nonsym}(b). As seen in the figure $\cC^k_1$ is maximized when $|a|=P, |b|=Q$ or $|a|=Q, |b|=P$ validating our conjecture.

\subsubsection{Case 2: $|a|=|b|$}

\begin{figure}[t]
\centering
\includegraphics[scale=0.4]{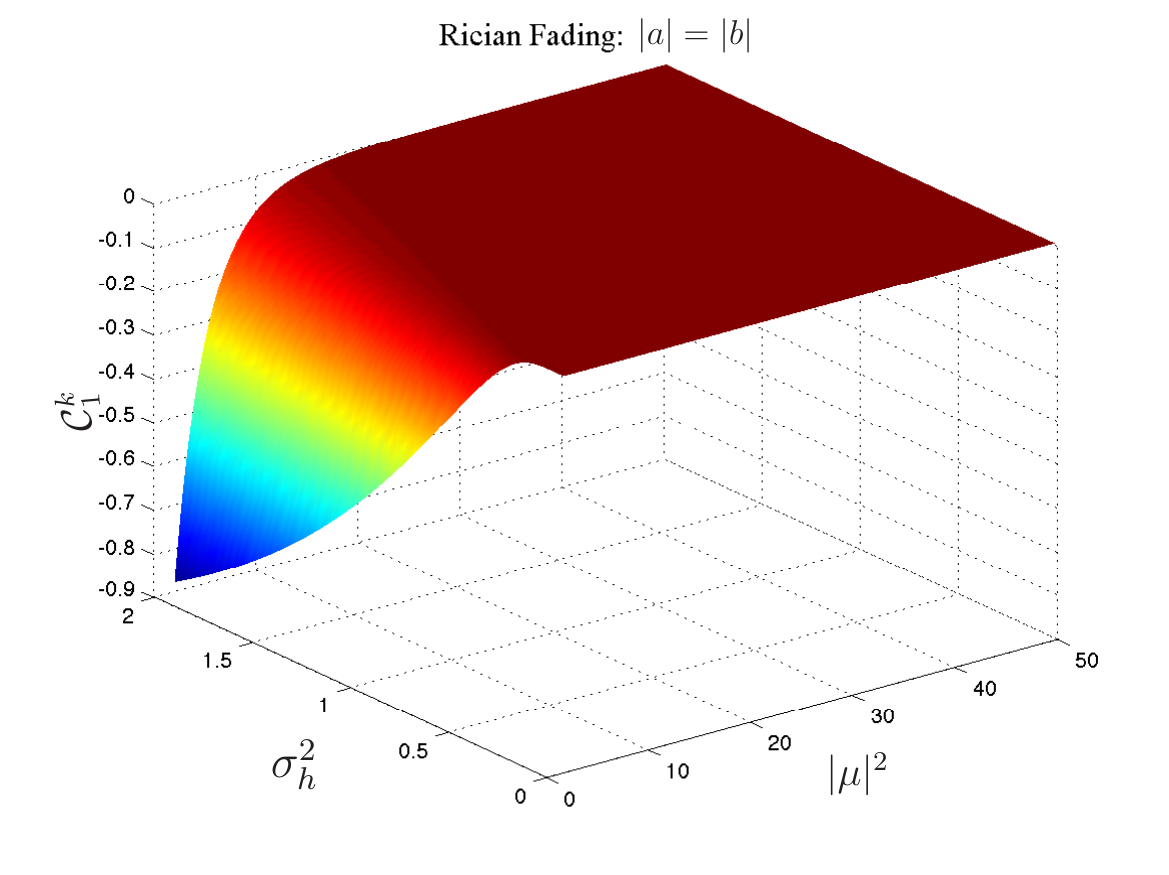}
\caption{The penalty term $\cC^k_1$ in Rician fading channels with $|a|=|b|$, as a function of the mean and the variance of the channel gain. $a=P=10$ and $b=-P=-10$.}
\label{fig:ric_sym}
\end{figure}

\ \ \ For $\kappa=0$, we have $\sigma_a^2=\sigma_b^2$, i.e., $|a|=|b|$. Accordingly from (\ref{eq:ric_perf2}) we write $\zeta_a = -s^2 -\frac{2}{\sigma_a^2}\gamma_a$, $\zeta_b = -s^2 -\frac{2}{\sigma_a^2}\gamma_b$ where similar to Section \ref{sec:awgn_per} we define $s \triangleq \frac{|a-b||\mu|}{\sigma_a}$, $\gamma_a \triangleq \Re\{\tilde{z}_a^* (a-b)\mu\}$ and $\gamma_b \triangleq \Re\{\tilde{z}_b^* (b-a)\mu\}$. Defining standard Gaussian random variables $\nu_a \triangleq \frac{\sqrt{2}}{|a-b||\mu|\sigma_a} \gamma_a$ and $\nu_b \triangleq \frac{\sqrt{2}}{|a-b||\mu|\sigma_a} \gamma_b$, analogous to the AWGN case, we have $\zeta_a = -s^2-\sqrt{2}s\nu_a$ and $\zeta_b = -s^2-\sqrt{2}s\nu_b$. Therefore, from (\ref{eq:ric_perf1}) $\cC^k_i$ is given by (\ref{eq:awgn_perf2}). Accordingly, Lemma \ref{lem:aw} applies here in the case of $|a|=|b|$ under Rician channels. This case is analogous to the AWGN case since the received signal $z^k_n$ has the same variance, but different means when $x^k_n=a$ and $x^k_n=b$. Consequently, antipodal signaling is optimal.
In Fig. \ref{fig:ric_sym}, $\cC^k_1$ is plotted as a function of the channel gain parameters $|\mu|^2$ and $\sigma_h^2$.  It is seen that $\cC^k_1$ is increasing in $|\mu|^2$ and decreasing in $\sigma_h^2$ when antipodal signaling is used, which corroborates Lemma \ref{lem:aw} since $s$ is increasing in $|\mu|^2$ and decreasing in $\sigma_h^2$.

\begin{figure}[t]
\centering
\includegraphics[scale=0.4]{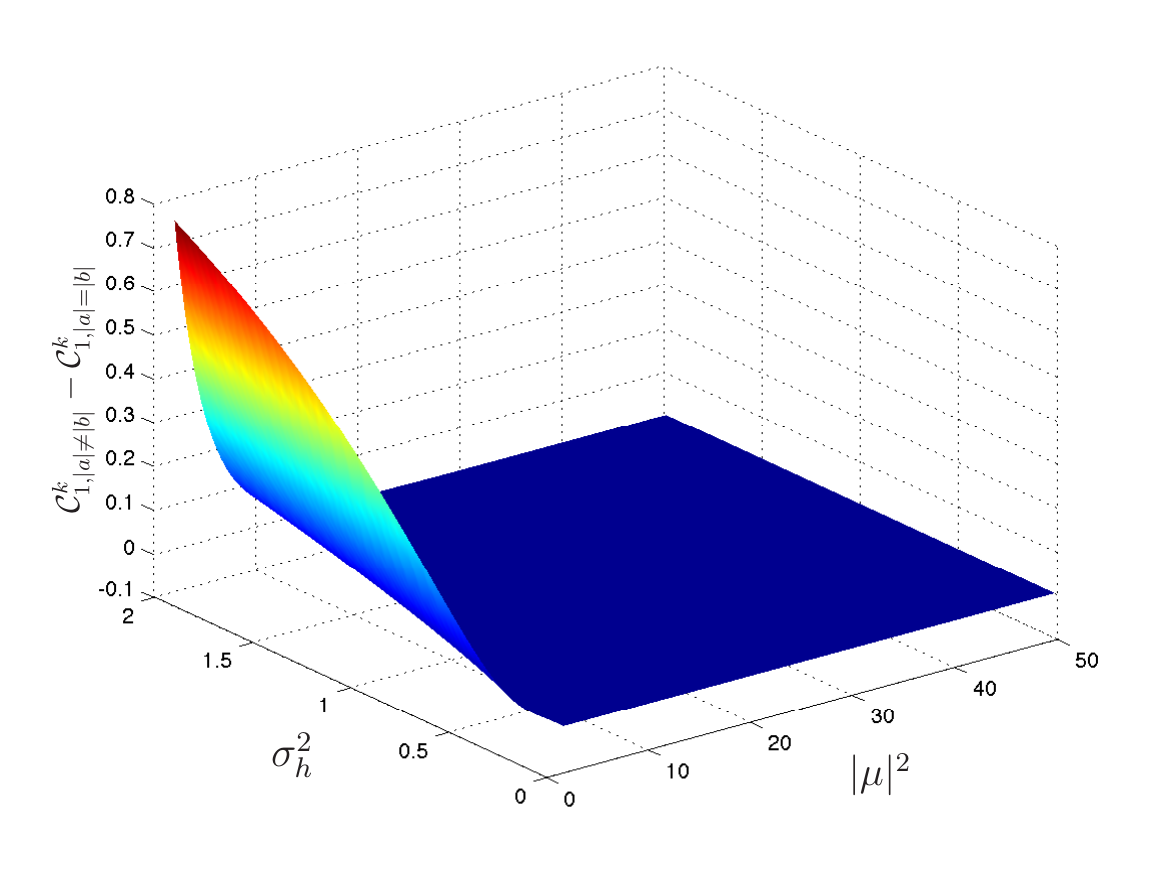}
\caption{$\cC^k_{1,|a|\neq|b|}-\cC^k_{1,|a|=|b|}$ in Rician fading channels as a function of $|\mu|^2$ and $\sigma_h^2$, $P=10, Q=1$.}
\label{fig:ric_comp}
\end{figure}

In Fig. \ref{fig:ric_comp}, the difference $\cC^k_{1,|a|\neq|b|}-\cC^k_{1,|a|=|b|}$ is plotted as a function of $|\mu|^2$ and $\sigma_h^2$. For $|a|=|b|$ antipodal signaling is employed; and for $|a|\neq |b|$, OOK-like signaling is employed. It is seen that the OOK-like signaling is much better than antipodal signaling when the mean is low and the variance is high. Although not visible in Fig. \ref{fig:ric_comp}, antipodal signaling is only slightly better than OOK-like signaling when the mean is high and the variance is low.

\section{Discussions}
\label{sec:Unrel_det}

\begin{figure}[!ht]
\centering
\includegraphics[scale=0.4]{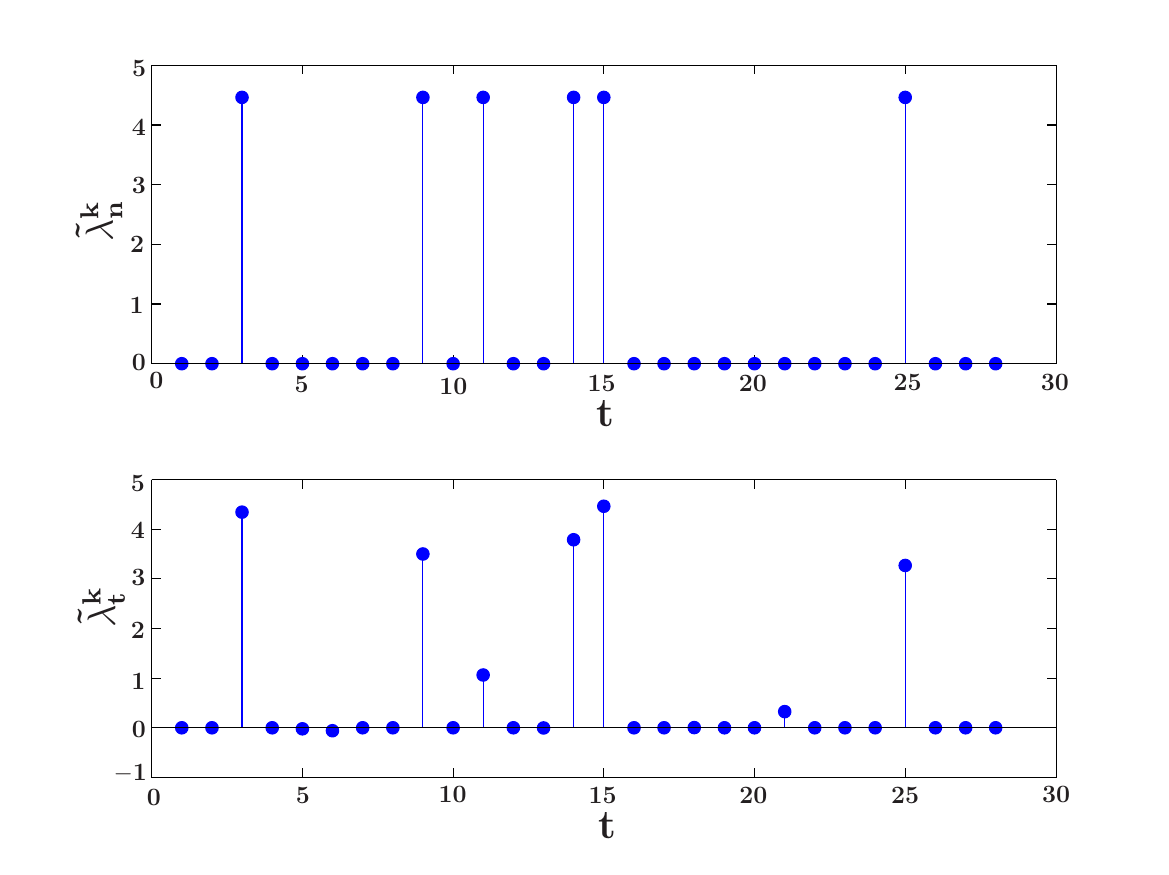}
\caption{Realizations of the LLRs $\tilde{\lambda}_n^k$ and $\tilde{\lambda}_t^k$ computed at the FC under reliable and unreliable detection of the sampling times, respectively.}
\label{fig:unr1}
\end{figure}

Considering the unreliable detection of the sampling times under continuous channels, we should ideally integrate this uncertainty into the fusion rule of the FC. In other words, at the FC the LLR $\tilde{\lambda}_t^k$ of the received signal $z_t^k$ should be computed at each time instant $t$ if the sampling time of the $k$-th sensor cannot be reliably detected. In the LLR computations in (\ref{eq:LLR_awgn}) and (\ref{eq:LLR_ray}) the prior probabilities $\Pro_i(x_n^k=a)$ and $\Pro_i(x_n^k=b)$ are used. These probabilities are conditioned on the sampling time $t_n^k$. Here, we need the unconditioned prior probabilities of the signal $x_t^k$ which at each time $t$ takes a value of $a$ or $b$ or $0$, i.e,
\be
    x_t^k = \left\{ \begin{array}{ll} a & \text{if} \ \ \ L_t^k-L_{t_{n-1}^k}^k \geq \Delta \\
                                    b & \text{if} \ \ \ L_t^k-L_{t_{n-1}^k}^k \leq -\Delta \\
                                    0 & \text{if} \ \ \ L_t^k-L_{t_{n-1}^k}^k \in (-\Delta,\Delta). \end{array} \right.
\ee
As before, the received signal at time $t$ is $z_t^k = h_t^k x_t^k + w_t^k$. Then, the LLR $\tilde{\lambda}_t^k$ of $z_t^k$ is given by
\be
\tilde{\lambda}_t^k = \log \frac{(1-\beta_k) \Pro_{s,1}^k p(z_t^k|x_t^k=a) + \beta_k \Pro_{s,1}^k p(z_t^k|x_t^k=b) + (1-\Pro_{s,1}^k) p(z_t^k|x_t^k=0)} {\alpha_k \Pro_{s,0}^k p(z_t^k|x_t^k=a) + (1-\alpha_k) \Pro_{s,0}^k p(z_t^k|x_t^k=b) + (1-\Pro_{s,0}^k) p(z_t^k|x_t^k=0)}
\ee
where $\Pro_{s,i}^k$ is the probability that the FC receives a signal from sensor $k$ under $\Hyp_i$. Since the FC has no prior information on the sampling times of the sensors, this probability can be shown to be $\frac{1}{\Exp_i[\tau_1^k]}$, where $\Exp_i[\tau_1^k]$ is the average intersampling (communication) interval for sensor $k$ under $\Hyp_i,~i=0,1$. For instance, under AWGN channels [cf. \eqref{eq:LLR_awgn}] by defining $c_t^k \triangleq \frac{|z_t^k-h_k a|^2}{\sigma_k^2}$, $d_t^k \triangleq \frac{|z_t^k-h_k b|^2}{\sigma_k^2}$, and $g_t^k \triangleq \frac{|z_t^k|^2}{\sigma_k^2}$ we have
\be
\label{eq:unr1}
\tilde{\lambda}_t^k = \log \frac{(1-\beta_k) \Pro_{s,1}^k e^{-c_t^k} + \beta_k \Pro_{s,1}^k e^{-d_t^k} + (1-\Pro_{s,1}^k) e^{-g_t^k}} {\alpha_k \Pro_{s,0}^k e^{-c_t^k} + (1-\alpha_k) \Pro_{s,0}^k e^{-d_t^k} + (1-\Pro_{s,0}^k) e^{-g_t^k}}.
\ee
Under fading channels $\tilde{\lambda}_t^k$ is computed similarly. Realizations of $\tilde{\lambda}_t^k$ of \eqref{eq:unr1} and $\tilde{\lambda}_n^k$ of \eqref{eq:LLR_awgn} are shown in Fig.\,\ref{fig:unr1} where $P=10$ is used.

Note that in this case, $\{\tilde{\lambda}_t^k\}$ are i.i.d. across time, and so are $\{\tilde{\lambda}_t\}$ where $\tilde{\lambda}_t \triangleq \sum_{k=1}^K \tilde{\lambda}_t^k$ is the global LLR at time $t$. Hence, from Wald's identity, similar to Theorem \ref{thm:id2} we can write $\Exp_1[\cT]=\frac{\Exp_1[\sum_{t=1}^T \tilde{\lambda}_t]}{\Exp_1[\tilde{\lambda}_t]}=\frac{|\log \alpha|}{\Exp_1[\tilde{\lambda}_t]}+O(1)$. Therefore, we again need to maximize the KL information $\Exp_1[\tilde{\lambda}_t^k]$ (resp. $-\Exp_0[\tilde{\lambda}_t^k]$) in order to minimize the average delay $\Exp_1[\cT]$ (resp. $\Exp_0[\cT]$). However, analyzing this expectation is now much more involved than analyzing \eqref{eq:awgn_perf1}. On the other hand, in practice we need to ensure reliable detection of the sampling times by using high enough signaling levels $P$ and $Q$. Then, the average delay performance of this unreliable detection scheme becomes identical to that of the reliable detection scheme analyzed in Section \ref{sec:Per_noi}.

As an alternative approach, in the unreliable detection case one can follow a two-step procedure to mimic the reliable detection case. Since it is known that most of the computed LLRs $\{\tilde{\lambda}_t^k\}$ are uninformative that correspond to the no message case, a simple thresholding operation can be applied to update the LLR only when it is informative. The thresholding step is in fact a Neyman-Pearson test between the presence and absence of a message signal. The threshold can be adjusted to control the false alarm and misdetection probabilities. Setting the threshold appropriately we can obtain a negligible false alarm probability, leaving us with the misdetection probability. Note that such a test would turn a continuous channel into a BEC with erasure probability, $\tilde{\epsilon}_k$, equal to the misdetection probability. Recall from Section \ref{sec:bec_alg} that under BEC $\tilde{\lambda}_n^k$ is the same as in the ideal channel case which corresponds to the reliable detection case here.
Thus, if an LLR survives after thresholding, in the second step it is recomputed as in the channel-aware fusion rules obtained in Sections \ref{sec:awgn_alg}, \ref{sec:ray_alg} and \ref{sec:ric_alg}. Moreover, the KL information in (\ref{eq:awgn_perf1}), (\ref{eq:ray_perf1}) and (\ref{eq:ric_perf1}) will only be scaled by $(1-\tilde{\epsilon}_k)$ as shown in (\ref{eq:bec_perf1}). Consequently, the results obtained in Sections \ref{sec:awgn_per}, \ref{sec:ray_per}, and \ref{sec:ric_per} are also valid in this approach to the unreliable detection case.

\section{Simulation Results}
\label{sec:sim}

In this section, we provide simulation results to illustrate the performance of the channel-aware distributed detection schemes based on level-triggered sampling. Assume there are two sensors collaborating with an FC. At each time $t$, each sensor makes a local observation $y_t^k= s + v_t^k,~k=1,2$, with $v_t^k \sim \cN_c(0,1)$,  $s=1$ under $\Hyp_1$, and $s=0$ under $\Hyp_0$. Hence, the LLR, $l_t^k$, of $y_t^k$ is computed as $l_t^k=2\Re\{y_t^k\}-1$.

Each sensor on average samples and transmits $1$ bit to the FC once every four samples they observe, i.e., $T=4$. And the local threshold $\Delta$ is determined to meet this average sampling interval. It has been shown in \cite[Section IV-A]{Yilmaz11} that one can use the equation $\Delta \tanh(\frac{\Delta}{2}) = T \frac{I_i(1)}{K}$ to find $\Delta$.
Then, using the $\Delta$ value the local error probabilities $\alpha_k$ and $\beta_k$ are computed offline for each sensor.
From Lemma \ref{lem:id} and Proposition \ref{pro:noi2}, we have $|\log \alpha|-\Theta \leq A,\tilde{A} \leq |\log \alpha|$ and $|\log \beta|-\Theta \leq B,\tilde{B} \leq |\log \beta|$ where $\Theta$ is the largest received LLR magnitude. Hence, we can set the global thresholds $\tilde{A}$ and $\tilde{B}$ to their upper bounds $|\log \alpha|$ and $|\log \beta|$ respectively to meet the constraints $\Pro_0(\delta_{\tilde{\cT}}=\Hyp_1)\leq\alpha$ and $\Pro_1(\delta_{\tilde{\cT}}=\Hyp_0)\leq\beta$. To achieve the equalities, $\tilde{A}$ (resp. $\tilde{B}$) should be found via simulations within the interval $[|\log \alpha|-\Theta,|\log \alpha|]$ (resp. $[|\log \beta|-\Theta,|\log \beta|]$). Note also that $A,\tilde{A}\sim|\log \alpha|$ and $B,\tilde{B}\sim|\log \beta|$ as $\alpha,\beta\to0$.

We compare our channel-aware designs with the conventional approach where the FC first decides on the received data bit and then uses it to update the test statistic.
Under BEC and BSC, since the received signal is already binary, in the conventional approach the FC simply treats the channel as ideal. On the other hand, under AWGN, Rayleigh fading, and Rician fading channels, in the conventional approach the FC first demodulates the received bit by using the following maximum-likelihood (ML) decision rules
\begin{align}
\text{AWGN :} & \max_{x^k_n \in \{a, b\}} \Re\{(z^k_n)^* h_k x^k_n\}, \nn\\
\text{Rayleigh :} & \max_{x^k_n \in \{a, b\}} \frac{\exp(-\frac{|z^k_n|^2}{2(|x^k_n|^2\sigma_h^2+\sigma^2)})}{|x^k_n|^2\sigma_h^2+\sigma^2}, \nn\\
\text{Rician :} & \max_{x^k_n \in \{a, b\}} \frac{\exp(-\frac{|z^k_n-x^k_n \mu|^2}{2(|x^k_n|^2\sigma_h^2+\sigma^2)})}{|x^k_n|^2\sigma_h^2+\sigma^2}. \nn
\end{align}
Then, it updates the test statistic either by treating the channel as ideal, i.e., using (\ref{eq:ideal2}), (note that this approach cannot guarantee to satisfy the target error probabilities since its performance highly depends on the performance of the receiver block) or more reasonably by treating the channel as a BSC assuming the error rate of the ML receiver is known, i.e., using (\ref{eq:bsc2}).

\subsection{Error Performance}

\begin{figure}[tb]
\centering
\includegraphics[scale=0.4]{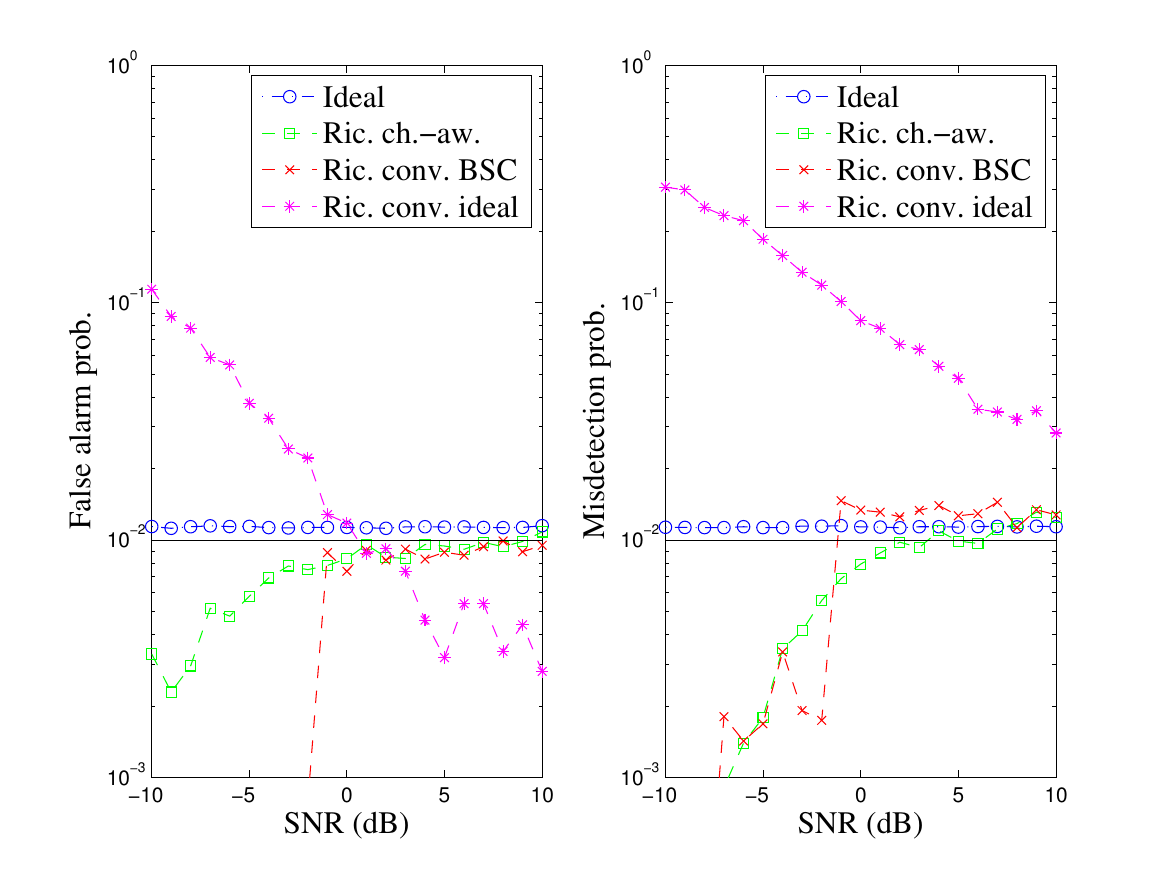}
\caption{Error performance comparison between the proposed channel-aware approach and the conventional methods.}
\label{fig:err}
\end{figure}

Firstly, we demonstrate that the channel-aware designs presented in this paper can meet the target error performance in noisy channels..
We set $\alpha=\beta=10^{-2}$, $\tilde{A}=|\log \alpha|$, $\tilde{B}=|\log \beta|$, $\Exp[|h^k_n|^2]=1$ (i.e., $|h_k|^2=1, \forall k$ for AWGN channels; $\sigma_{h,k}^2=1, \forall k$ for Rayleigh fading channels; and $\sigma_{h,k}^2=0.5, |\mu_k|^2=0.5, \forall k$ for Rician fading channels). We define SNR$\triangleq \frac{\Exp[|h^k_n|^2]}{\sigma_k^2}$.
As an example, in Fig. \ref{fig:err} we show the actual error performances in Rician fading channels for both the proposed channel-aware approach and the conventional methods. The error performance under ideal channels is also shown. It is seen that the channel-aware method and the conventional method treating the channel as BSC can always meet the specified error bounds under different channel conditions. In fact, they achieve even smaller error probabilities under bad channel conditions, i.e., low SNR or high $\epsilon$,  since they update the test statistics even more cautiously with smaller increments. However, the conventional approach that treats the channel as ideal is vulnerable to noisy channels. Its error performances are far away from the bounds especially at low SNR. Similar results are observed for the other noisy  channel types.

\subsection{Detection Delay Performance }

We now show the actual  decision delay performance of the proposed channel-aware approach  as a function of the achieved error rates. In this subsection, different from the previous one we do not determine the thresholds $\tilde{A}$ and $\tilde{B}$ for the given error probability bounds. But rather, for a specific set of $\tilde{A}$ and $\tilde{B}$ values we simulate the schemes to obtain their operating characteristics, i.e., the average decision delay and error probabilities.
For fair comparisons we set the channel error probabilities of discrete channels, i.e., BEC and BSC, to $\epsilon_k=0.1, \forall k$; and set SNR$=0$dB for all continuous channel types.
\begin{figure}[tb]
\centering
\includegraphics[scale=0.4]{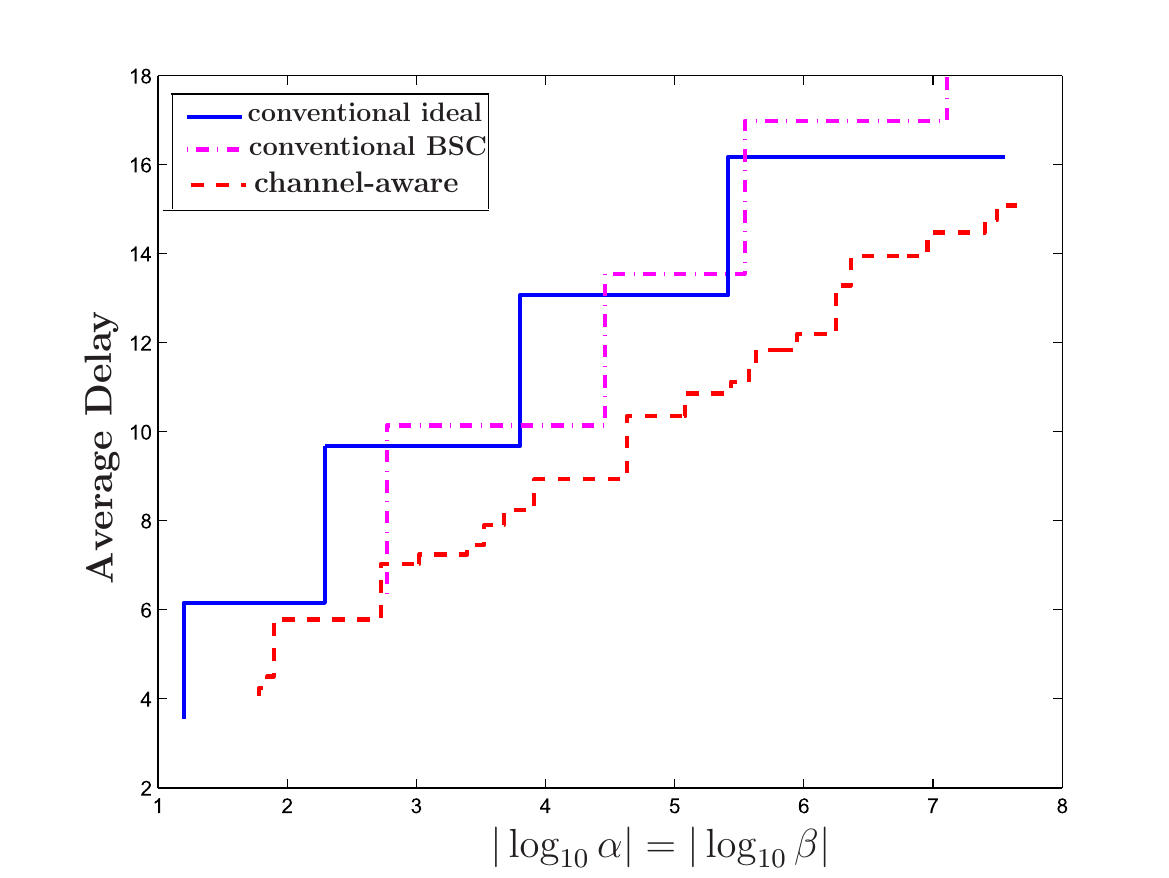}
\caption{The average decision delay vs. the achieved error rates in Rician fading channels.}
\label{fig:ric_class_chaw}
\end{figure}
Fig. \ref{fig:ric_class_chaw} compares the channel-aware scheme to the two conventional schemes assuming ideal channels and BSC, respectively, after bit recovery under Rician fading channels.
The average decision delay of the channel-aware scheme is significantly lower than those of the conventional schemes. Moreover, the channel-aware scheme provides more achievable error probabilities than the conventional schemes, since the step sizes are much finer for the channel-aware scheme. The discrete nature of the average decision delay curve is due to having finite number of values to update the test statistic at the FC. This phenomenon was explained in detail in \cite{Yilmaz11}. The conventional schemes have only two possible update values that are given in (\ref{eq:ideal2}) and (\ref{eq:bsc2}), whereas the channel-aware scheme uses a continuum of values to update its LLR sum as given in (\ref{eq:LLR_ray}). Similar results can be obtained for the other channel types.

\begin{figure}[tb]
\centering
\includegraphics[scale=0.4]{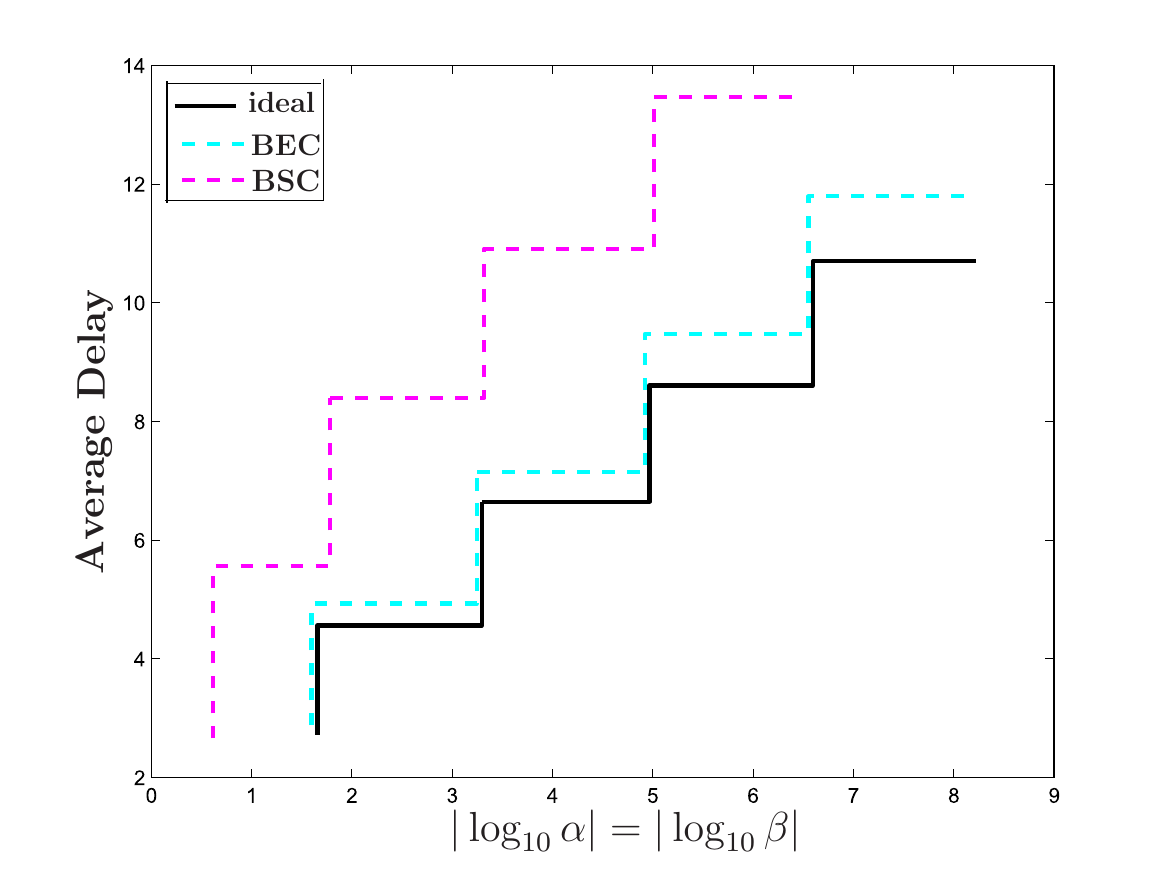}
\caption{The average decision delay as a function of the achieved error rate under different discrete channels.}
\label{fig:id_chaw_disc}
\end{figure}

\begin{figure}[tb]
\centering
\includegraphics[scale=0.4]{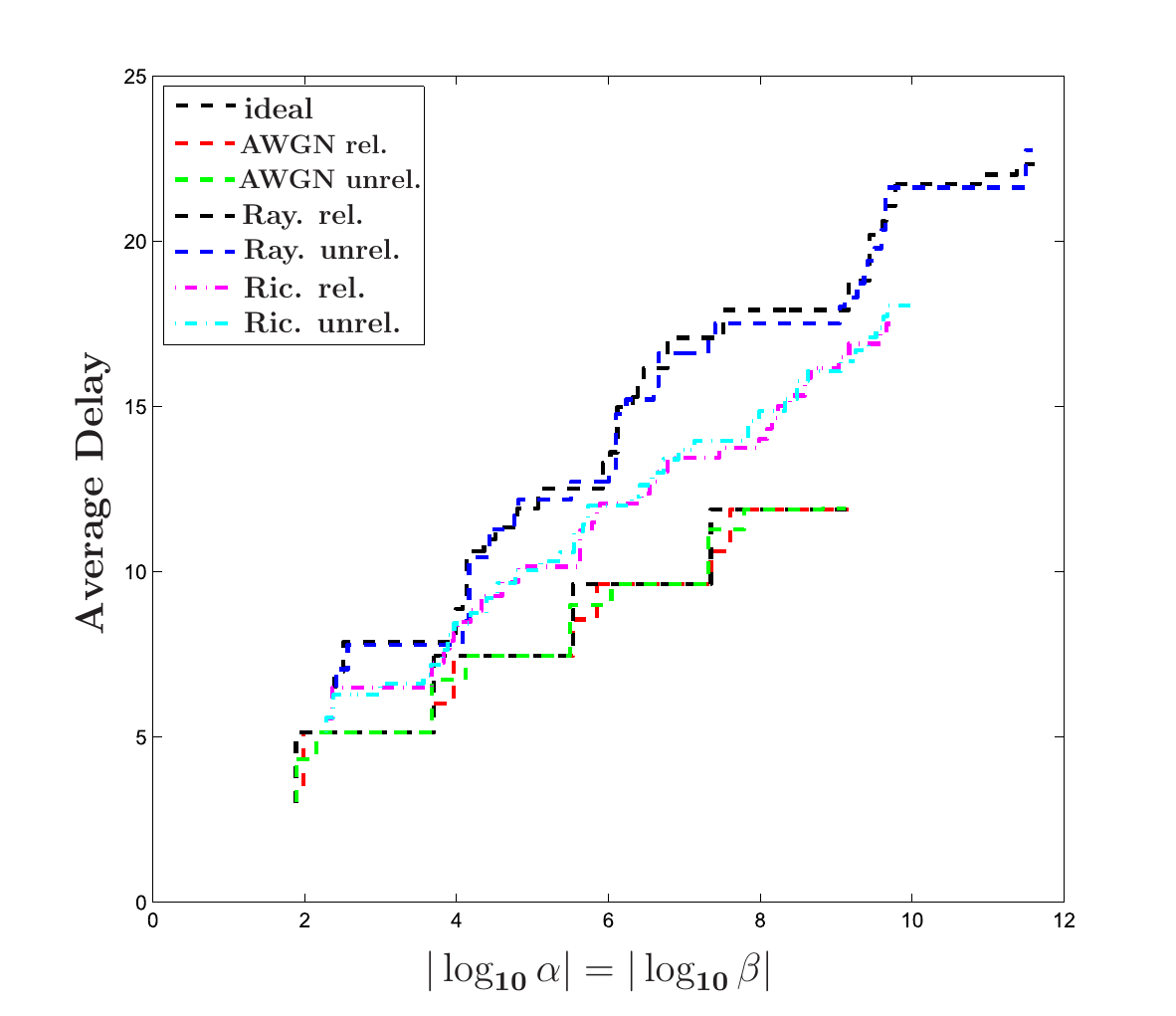}
\caption{The average decision delay as a function of the achieved error rate under different continuous channels. }
\label{fig:id_chaw_cont}
\end{figure}
Next, we compare the decision delay performances of the channel-aware schemes under different channels. Fig. \ref{fig:id_chaw_disc} and Fig. \ref{fig:id_chaw_cont} show the results for the discrete channels and the continuous channels, respectively. It is seen that BEC has a superior performance than BSC. Note from Fig. \ref{fig:id_chaw_disc} that the step sizes are large and the number of achievable error probabilities is the same for all three cases since there are only two LLR update values [cf. (\ref{eq:ideal2})-(\ref{eq:bsc3})].  For each continuous channel type, the corresponding signaling scheme discussed in Sections V.C-E is used in the simulations. As expected the AWGN channel case has a much better performance than the fading channel cases since under AWGN, the channels are deterministic and known to the FC, whereas in fading cases the channels are random and only the statistics are known to the FC. Moreover, under Rayleigh fading, channels have zero mean increasing the uncertainty, hence this case has the worst performance among the continuous channel types. Finally we consider the fusion rule \eqref{eq:unr1} that takes into account the unreliable detection of the sampling times.
We use SNR$=0$ dB for all channels; $P=20$ under AWGN; $P=Q=20$ under Rician; and $P=100, Q=20$ under Rayleigh.
In Fig. \ref{fig:id_chaw_cont}, it is seen that the channel aware scheme has almost identical performances in the reliable and unreliable detection cases under all continuous channels.

\section{Conclusions}
\label{sec:conc}

We have developed and analyzed channel-aware distributed detection schemes based on level-triggered sampling. The sensors form local log-likelihood ratios (LLRs) based on their observations and sample their LLRs using the level-triggered sampling.
Upon sampling each sensor sends a single bit to the fusion center (FC).
The FC is equipped with the local error rates of all sensors and the statistics of the channels from all sensors. Upon receiving the bits from the sensors, the FC updates the global LLR and performs an SPRT. The fusion rules under different channel types are given. We have further provided an asymptotic analysis on the average decision delay for the proposed channel-aware scheme. We have shown that the asymptotic decision delay  is   characterized by a KL information
number, whose expressions under different channel types have been derived.  Based on the delay analysis, we have also identified appropriate signaling schemes under different channels for the sensors to transmit the 1-bit information. Numerical examples have demonstrated the advantages of the proposed channel-aware approach over the conventional methods.

\abovedisplayskip=0.15cm
\belowdisplayskip=0.15cm
\section*{Appendix: Proof of Lemma 2}
We will present the proof under $\Hyp_1$, and the proof under $\Hyp_0$ follows similarly. We need to  find the condition for $\frac {d {\cal C}_1^k} {ds}>0$. From (\ref{eq:awgn_perf2}), we have
\begin{multline}
\frac {d {\cal C}_1^k} {ds} = \Exp\left[ \frac{(\sqrt{2}\nu-2s) e^{-s^2+\sqrt{2}s\nu}(G-F^{-1})} {(1+Ge^{-s^2+\sqrt{2}s\nu})(1+F^{-1}e^{-s^2+\sqrt{2}s\nu})} + \frac{(-\sqrt{2}\nu-2s) e^{-s^2-\sqrt{2}s\nu}(1-FG)} {(1+G^{-1}e^{-s^2-\sqrt{2}s\nu})(1+Fe^{-s^2-\sqrt{2}s\nu})} \right]  \\
= (FG-1) \int_{-\infty}^{\infty} \left[ \frac{(\sqrt{2}\nu-2s) e^{-s^2+\sqrt{2}s\nu}} {F(1+Ge^{-s^2+\sqrt{2}s\nu})(1+F^{-1}e^{-s^2+\sqrt{2}s\nu})} - \right.\\
\left. \frac{(-\sqrt{2}\nu-2s) e^{-s^2-\sqrt{2}s\nu}} {(1+G^{-1}e^{-s^2-\sqrt{2}s\nu})(1+Fe^{-s^2-\sqrt{2}s\nu})}  \right] \frac{e^{-\frac{\nu^2}{2}}}{\sqrt{2\pi}} {\rm d}\nu \\
= \frac{FG-1}{\sqrt{2\pi}} \int_{-\infty}^{\infty} \left[ \frac{(\sqrt{2}\nu-2s) e^{-\left( \frac{\nu}{\sqrt{2}}-s \right)^2}} {F(1+Ge^{-s^2+\sqrt{2}s\nu})(1+F^{-1}e^{-s^2+\sqrt{2}s\nu})} + \frac{(\sqrt{2}\nu+2s) e^{-\left( \frac{\nu}{\sqrt{2}}+s \right)^2}} {(1+G^{-1}e^{-s^2-\sqrt{2}s\nu})(1+Fe^{-s^2-\sqrt{2}s\nu})}  \right] {\rm d}\nu
\label{app:aw1}
\end{multline}
If we choose $\Delta>0$, then we will have $\alpha_k+\beta_k<1$ which in turn yields $FG>1$, but here we will reasonably assume that $\alpha_k,\beta_k<0.5$ and accordingly $F,G>1$. Therefore, it is clear that in order to conclude the proof we need to show that the integral in (\ref{app:aw1}) is positive. Define $r_1 \triangleq \frac{\nu}{\sqrt{2}}-s$ and $r_2 \triangleq \frac{\nu}{\sqrt{2}}+s$, then we need to show the following inequality
\begin{multline}
\label{app:aw2}
\int_0^{\infty} \left[ \frac{2r_1 e^{-r_1^2}} {F(1+Ge^{s^2+2s r_1})(1+F^{-1}e^{s^2+2s r_1})} - \frac{2r_1 e^{-r_1^2}} {F(1+Ge^{s^2-2s r_1})(1+F^{-1}e^{s^2-2s r_1})} \right] {\rm d}r_1 > \\
- \int_0^{\infty} \left[ \frac{2r_2 e^{-r_2^2}} {(1+G^{-1}e^{s^2-2s r_2})(1+Fe^{s^2-2s r_2})} - \frac{2r_2 e^{-r_2^2}} {(1+G^{-1}e^{s^2+2s r_2})(1+Fe^{s^2+2s r_2})} \right] {\rm d} r_2.
\end{multline}
Note that (\ref{app:aw2}) holds if the following inequality holds,
\begin{multline}
\frac{1}{F(1+Ge^{s^2+2s r})(1+F^{-1}e^{s^2+2s r})} - \frac{1}{F(1+Ge^{s^2-2s r})(1+F^{-1}e^{s^2-2s r})} > \\
-\frac{1}{(1+G^{-1}e^{s^2-2s r})(1+Fe^{s^2-2s r})} + \frac{1}{(1+G^{-1}e^{s^2+2s r})(1+Fe^{s^2+2s r})}.
\end{multline}
Thus, after rearranging terms it is sufficient to show that
\begin{multline}
\frac{(G^{-1}F-G) e^{s2^2+4sr} + (FG+1)(G^{-1}-1) e^{s^2+2sr} + (1-F)}  {(1+Ge^{s^2+2s r})(1+F^{-1}e^{s^2+2s r})(1+G^{-1}e^{s^2+2s r})(1+Fe^{s^2+2s r})} > \\
\frac{(G^{-1}F-G) e^{2s^2-4sr} + (FG+1)(G^{-1}-1) e^{s^2-2sr} + (1-F)}  {(1+Ge^{s^2-2s r})(1+F^{-1}e^{s^2-2s r})(1+G^{-1}e^{s^2-2s r})(1+Fe^{s^2-2s r})}.
\label{app:aw3}
\end{multline}
Define $p\triangleq s^2+2sr$, $q\triangleq s^2-2sr$, $C_1 \triangleq G-\frac{F}{G}$, $C_2 \triangleq (FG+1)(1-\frac{1}{G})$, $C_3 \triangleq F+F^{-1}+G+G^{-1}$, $C_4 \triangleq 2+FG+\frac{F}{G}+\frac{G}{F}+\frac{1}{FG}$. Multiplying both sides with $-1$, and rearranging terms we can rewrite (\ref{app:aw3}) as follows
\begin{multline}
(C_1e^{2p}+C_2e^p+F-1) (e^{4q}+C_3e^{3q}+C_4e^{2q}+C_3e^q+1) < \\
(C_1e^{2q}+C_2e^q+F-1) (e^{4p}+C_3e^{3p}+C_4e^{2p}+C_3e^p+1).
\label{app:aw5}
\end{multline}
After some manipulations, we obtain the following inequality
\begin{multline}
C_1C_3(e^{2p+q}-e^{p+2q}) + C_1(e^{2p}-e^{2q}) + C_2(e^p-e^q) < \\
C_1(e^{4p+2q}-e^{2p+4q}) + C_2(e^{4p+q}-e^{p+4q}) + \\
(F-1)(e^{4p}-e^{4q}) + C_1C_3(e^{3p+2q}-e^{2p+3q}) + \\
C_2C_3(e^{3p+q}-e^{p+3q}) + C_3(F-1)(e^{3p}-e^{3q}) + \\
C_2C_4(e^{2p+q}-e^{p+2q}) + C_4(F-1)(e^{2p}-e^{2q}) + C_3(F-1)(e^p-e^q).
\end{multline}
Finally, noting that $p>q$ (since $s>0, r>0$) if we cancel the common term $e^p-e^q$, then the inequality that we need to verify becomes the following
\begin{multline}
C_1C_3e^{p+q} + C_1(e^p+e^q) + C_2 < \\
C_1C_3e^{2p+2q} + C_1e^{2p+2q}(e^p+e^q) + C_2e^{p+q}(e^{2p}+e^{p+q}+e^{2q}) + \\
(F-1)(e^{2p}+e^{2q})(e^p+e^q) + C_2C_3e^{p+q}(e^p+e^q) + \\
C_3(F-1)(e^{2p}+e^{p+q}+e^{2q}+1) + C_2C_4e^{p+q} + C_4(F-1)(e^p+e^q).
\label{app:aw6}
\end{multline}

Now assuming that $C_1\geq0$, i.e., $G^2\geq F$, it is straightforward to verify the inequality in (\ref{app:aw6}). Since we have $p+q=s^2>0$, we also have $e^{p+q}<(e^{p+q})^2$, $e^p+e^q<e^{2p+2q}(e^p+e^q)$, and $e^{p+q}(e^{2p}+e^{p+q}+e^{2q})>1$. Note also that the last five terms on the right hand side of (\ref{app:aw6}) are positive due to having $F>1, C_1>0, C_2>0, C_3>0, C_4>0$. Hence, $\cC^k_1$ is increasing in $s$ for all $k$ when $G^2\geq F$. Similarly we can show that  $\cC^k_0$ is increasing in $s$ for all $k$ when $F^2\geq G$.

%\bibliographystyle{IEEEbib}
%\bibliography{IEEEabrv,Yasin}

\begin{thebibliography}{99}
\bibitem{Tenney81}
R.R.~Tenney, and N.R.~Sandell,
\newblock ``Detection with distributed sensors,''
\newblock {\em IEEE Trans. Aero. Electron. Syst.}, vol. 17, no. 4, pp. 501-510, July 1981.

\bibitem{Chair86}
Z.~Chair, and P.K.~Varshney,
\newblock ``Optimal data fusion in multiple sensor detection systems,''
\newblock {\em IEEE Trans. Aero. Electron. Syst.}, vol. 22, no. 1, pp. 98-101, Jan. 1986.

\bibitem{Thomo87}
S.C.A~Thomopoulos, R.~Viswanathan, and D.C.~Bougoulias,
\newblock ``Optimal decision fusion in multiple sensor systems,"
\newblock {\em  IEEE Trans. Aero. Electron. Syst.}, vol. 23, no. 5,
  pp. 644-653, Sept. 1987.

\bibitem{Tsitsiklis88}
J.~Tsitsiklis,
\newblock ``Decentralized detection by a large number of sensors,''
\newblock {\em Mathematics of Control, Signals, and Systems}, pp. 167-182, 1988.

\bibitem{Aalo89}
V.~Aalo, and R.~Viswanathou,
\newblock ``On distributed detection with correlated sensors: two examples,"
\newblock {\em  IEEE Trans. Aero. Electron. Syst.}, vol. 25, no. 3,
 pp. 414-421, May 1989.

\bibitem{Willett00}
P.~Willett, P.F.~Swaszek, and R.S.~Blum,
\newblock ``The good, bad and ugly: distributed detection of a known signal in dependent
gaussian noise,''
\newblock {\em IEEE Trans. Sig. Proc.}, vol. 48,
  no. 12, pp. 3266-3279, Dec. 2000.

\ignore{
\bibitem{Teneketzis87}
D.~Teneketzis, and Y.-C.~Ho,
\newblock ``The decentralized Wald problem,''
\newblock {\em  Information and Computation}, vol. 73, no. 1,
  pp. 23-44, 1987.
}

\bibitem{Veer93}
V.V.~Veeravalli, T.~Basar, and H.V.~Poor, 
\newblock ``Decentralized sequential
detection with a fusion center performing the sequential test," 
\newblock {\em IEEE Trans. Inform. Theory}, vol. 39, no. 2, pp. 433Ð442, Mar. 1993.

\bibitem{Mei08}
Y.~Mei,
\newblock ``Asymptotic optimality theory for sequential hypothesis
testing in sensor networks," 
\newblock {\em IEEE Trans. Inform. Theory}, vol. 54, no. 5,
pp. 2072Ð2089, May 2008.

\bibitem{Chaud09}
S.~Chaudhari, V.~Koivunen, and H.V.~Poor,
\newblock ``Autocorrelation-based decentralized sequential detection of OFDM signals in
cognitive radios,"
\newblock {\em IEEE Trans. Sig. Proc.}, vol. 57, no. 7, pp. 2690-2700, July 2009.

\bibitem{Hussain94}
A.M.~Hussain,
\newblock ``Multisensor distributed sequential detection,"
{\em  IEEE Trans. Aero. Electron. Syst.}, vol. 30, no. 3,
 pp. 698-708, July 1994.

\bibitem{Fellouris11}
G.~Fellouris, and G.V.~Moustakides,
\newblock ``Decentralized sequential hypothesis testing using asynchronous
  communication,''
\newblock {\em IEEE Trans. Inform. Theory}, vol. 57, no. 1, pp.
  534-548, Jan. 2011.

\bibitem{Yilmaz11}
Y.~Yilmaz, G.V.~Moustakides, and X.~Wang,
\newblock ``Cooperative sequential spectrum sensing based on level-triggered sampling,''
\newblock {\em IEEE Trans. Sig. Proc.}, vol. 60, no. 9, pp.
  4509-4524, Sep. 2012.

\bibitem{Wald48}
A.~Wald and J.~Wolfowitz,
\newblock ``Optimum character of the sequential probability ratio test,''
\newblock {\em Ann. Math. Stat.}, vol. 19, pp. 326-329, 1948.

\bibitem{Poor}
H.V. Poor, {\em An Introduction to Signal Detection and Estimation}, 2nd edition, Springer, New York, NY, 1994.

\bibitem{Warren89}
D.J.~Warren, and P.K.~Willett,
\newblock ``Optimal decentralized detection for conditionally independent sensors,''
\newblock in Proc. 1989 Amer. Control Conf., pp. 1326-1329, June 1989.

\bibitem{Chaud12}
S.~Chaudhari, J.~Lunden, V.~Koivunen, and H.V.~Poor,
\newblock ``Cooperative sensing with imperfect reporting channels: Hard
decisions or soft decisions?,"
\newblock {\em IEEE Trans. Sig. Proc.}, vol. 60, no. 1, pp. 18-28, Jan. 2012.

\bibitem{Chen06}
B.~Chen, L.~Tong, and P.K.~Varshney,
\newblock ``Channel-aware distributed detection in wireless sensor networks,''
\newblock {\em IEEE Sig. Proc. Mag.}, vol. 23, no. 4, pp. 16-26, July 2006.

\bibitem{Chamberland03}
J.-F.~Chamberland, and V.V.~Veeravalli,
\newblock ``Decentralized detection in sensor networks,''
\newblock {\em IEEE Trans. Sig. Proc.}, vol. 51, no. 2, pp. 407-416, Feb. 2003.

\bibitem{Liu06}
B.~Liu and B.~Chen,
\newblock ``Channel-optimized quantizers for decentralized detection in sensor networks,''
\newblock {\em IEEE Trans. Inform. Theory}, vol. 52, no. 7, pp.
  3349-3358, July 2006.

\bibitem{Chen04}
B.~Chen, R.~Jiang, T.~Kasetkasem, and P.K.~Varshney,
\newblock ``Channel aware decision fusion in wireless sensor networks,''
\newblock {\em IEEE Trans. Sig. Proc.}, vol. 52, no. 12, pp. 3454-3458, Dec. 2004.

\bibitem{Niu06}
R.~Niu, B.~Chen, and P.K.~Varshney,
\newblock ``Fusion of decisions transmitted over Rayleigh fading channels in wireless
sensor networks,''
\newblock {\em IEEE Trans. Sig. Proc.}, vol. 54, no. 3, pp. 1018-1027, Mar. 2006.

\bibitem{Bahceci05}
I.~Bahceci, G.~Al-Regib, and Y.~Altunbasak,
\newblock ``Serial distributed detection for wireless sensor networks,''
\newblock in Proc. 2005 IEEE Int'l Symp. Inform. Theory (ISIT'05), pp. 830-834, Sept. 2005.

\bibitem{Tepedelen11}
C.~Tepedelenlioglu, and S.~Dasarathan,
\newblock ``Distributed detection over Gaussian multiple access channels with constant modulus
signaling,''
\newblock {\em IEEE Trans. Sig. Proc.}, vol. 59, no. 6, pp. 2875-2886, June 2011.

\bibitem{Ahmadi11}
H.R.~Ahmadi, and A.~Vosoughi,
\newblock ``Distributed detection with adaptive topology and nonideal communication channels,''
\newblock {\em IEEE Trans. Sig. Proc.}, vol. 59, no. 6, pp. 2857-2874, June 2011.

\bibitem{Wald47}
A.~Wald,
\newblock {\em Sequential Analysis},
\newblock Wiley, New York, NY, 1947.

\bibitem{Ross96}
S.~Ross,
\newblock {\em Stochastic Processes},
\newblock Wiley, New York, NY, 1996.

\end{thebibliography}

\end{document}